\documentclass[preprint,aps,tightenlines,showpacs,superscriptaddress]{revtex4}
\usepackage[dvips]{graphicx}
\usepackage{dcolumn}
\usepackage{epsfig}
\usepackage{color}

\begin{document}
\title{Extraction of Electromagnetic 
Transition Form Factors for Nucleon Resonances within a
Dynamical Coupled-Channels Model}
\author{N. Suzuki}
\affiliation{Department of Physics, Osaka University, Toyonaka,
Osaka 560-0043, Japan}
\affiliation{ Excited Baryon Analysis Center (EBAC), Thomas Jefferson National
Accelerator Facility, Newport News, VA 23606, USA}
\author{T. Sato}
\affiliation{Department of Physics, Osaka University, Toyonaka,
Osaka 560-0043, Japan}
\affiliation{ Excited Baryon Analysis Center (EBAC), Thomas Jefferson National
Accelerator Facility, Newport News, VA 23606, USA}
\author{T.-S. H. Lee}
\affiliation{Physics Division, Argonne National Laboratory,
Argonne, IL 60439, USA}
\affiliation{ Excited Baryon Analysis Center (EBAC), Thomas Jefferson National
Accelerator Facility, Newport News, VA 23606, USA}

\begin{abstract}
We explain the application of a recently developed
 analytic continuation method to extract
the electromagnetic transition form factors for
the nucleon resonances ($N^*$)  within a dynamical
coupled-channel model of meson-baryon reactions.
Illustrative 
results of the obtained $N^*\rightarrow \gamma N$ transition
 form factors, defined at the resonance pole positions on the complex
energy plane,
for the well isolated  $P_{33}$ and $D_{13}$, and the complicated
$P_{11}$ resonances are presented. 
A formula has been
developed to give an unified representation of the effects due to
the first two $P_{11}$  poles, which are near
the $\pi\Delta$ threshold, but are on different Riemann sheets.
We also find that a simple formula, with
its parameters determined in the Laurent expansions
of  $\pi N \rightarrow \pi N$ and $\gamma N \rightarrow\pi N$ amplitudes,
can reproduce to a very large extent
the exact solutions of the considered model at energies near the real parts of
the extracted resonance positions.
We indicate  the differences between 
our results and those extracted from the approaches using the Breit-Wigner 
parametrization
of resonant amplitudes to fit the data.

\end{abstract}

\maketitle

\newpage

\section{Introduction}

The spectrum and form factors of excited nucleons
are fundamental quantities
for investigating the hadron structure within Quantum Chromodynamics (QCD).
The excited nucleons are unstable and couple strongly to
meson-baryon continuum states to form nucleon
resonances (called collectively as $N^*$) in
$\pi N$ and $\gamma N$ reactions.
It is well known that  resonances locate on the unphysical
sheets of complex energy plane and thus their properties can only be
extracted from the empirical partial-wave amplitudes by
analytic continuation.
Recently we have applied
an analytic continuation method developed in
Ref.\cite{ssl09} to extract $N^*$ pole positions\cite{sjklms10} from
$\pi N$ elastic scattering amplitudes determined in a fit\cite{jlms07}
(JLMS) within a 
dynamical
coupled channel
model\cite{msl07} (EBAC-DCC) of meson-baryon reactions.

The scattering amplitudes obtained from  a dynamical coupled-channels
model of meson-baryon reactions, such as the EBAC-DCC model as well as
the models developed in Refs.\cite{afnan,gross,sl96,ntuanl,juelich-0},
are not available in an analytic form. They  are obtained numerically by
solving coupled-channels integral equations with meson-exchange driving terms.
Thus, the predicted
amplitudes can only be analytically continued to complex energy plane
numerically with a careful account of the analytic structure
of the considered scattering equations. Obviously, the method depends on the
dynamical content of each model. For EBAC-DCC model, this has been developed in
Ref.\cite{ssl09} and established using several exactly soluble models. 
In this paper, we explain how this method is used to extract
$\gamma^* N\rightarrow N^*$ transition form factors from the multipole
amplitudes determined from extending
the JLMS analysis to investigate $\gamma N \rightarrow \pi N$\cite{jlmss08}
and $N(e,e'\pi)N$\cite{jklmss09} reactions.

The electromagnetic $\gamma^* N \rightarrow N^*$ transition form factors
give information
on the current and charge distributions  of $N^*$ and $N$.
It can be shown\cite{bohm,dalitz}  that 
a resonance state $|\psi^R_{N^*}>$ with a complex energy $M_R$
can be defined as  an   'eigenstate' of
Hamiltonian $H|\psi^R_{N^*}> = M_R |\psi^R_{N^*}>$
 with the outgoing boundary condition
for its asymptotic wave functions.
Therefore the $\gamma^* N\rightarrow N^*$ transition form factor
is defined by 
the current matrix element $<\psi^R_{N^*}| J_{em} |N>$
which can be extracted from
the residue $R_{\pi N, \gamma^* N}$ of electromagnetic pion production 
amplitudes at the resonance poles.  To extract
$R_{\pi N, \gamma^* N}$, we need to evaluate the on-shell matrix elements
of 
 $\gamma^* N \rightarrow \pi N$ amplitudes on the complex Riemann energy 
sheet. As will be discussed later,
the analytic structure of the considered coupled-channels equations
for getting these on-shell matrix elements
is rather complex and must be dealt with carefully. In particular, we
need to develop a formula to give an unified representation of
the first two $P_{11}$ resonances which are near the $\pi\Delta$
threshold, but are on different Riemann sheets.

To illustrate our approach, it is sufficient to only present results
for the well isolated resonances in $P_{33}$ and $D_{13}$ and the
complex $P_{11}$ partial waves. 
With
only three complex parameters determined in the Laurent expansion
of each partial-wave amplitude at resonance pole position, we present
a simple formula 
which can reproduce to a very large extent
the exact solutions of the considered model at energies near the real parts of
the extracted resonance positions. 
This finding agrees with what 
was reported in an analysis\cite{juelich} of $\pi N$ scattering
amplitude within the J\"{u}lich model\cite{juelich-0}. 
Here we show that this formula is also a good approximation for
 $\gamma N \rightarrow \pi N$ amplitudes. Despite that
this formula is similar to
that used in the
analysis\cite{gwu-vpi,maid,inna} using the Breit-Wigner parametrization
of resonant amplitudes to fit the data, we find no simple relation between
two approaches.

In section II, we will briefly review 
the analytic continuation method developed in Ref.\cite{ssl09} and
explain how it is applied to  evaluate the on-shell amplitudes
of $\pi N, \gamma^* N \rightarrow \pi N$ transitions.
Section III is devoted to explaining how the
determined 
 residues are used to extract 
the elasticity $\eta_{el}$ of $N^* \rightarrow \pi N$ decay and
the $\gamma^* N \rightarrow N^*$ transition
form factors at  resonance poles. 
The results for $P_{11}$, $P_{33}$, 
and $D_{13}$ nucleon resonances are presented in section IV.
A summary is given in section V.

\section{Analytic continuation method}

Within the  formulation\cite{msl07} for EBAC-DCC model,
the partial wave amplitudes of 
meson-baryon reactions
can be written 
as
\begin{eqnarray}
T_{\beta,\alpha}(p',p;E)
& = & t_{\beta,\alpha}(p',p;E)
   + t^R_{\beta,\alpha}(p',p;E) \,,
\label{eq:fullt}
\end{eqnarray}
where $\alpha, \beta$ represent the  meson-baryon (MB)
 states $\gamma N$, $\pi N,\eta N,\rho N, \sigma N, \pi\Delta$, and
\begin{eqnarray}
t^R_{\beta,\alpha}(p',p;E) &=& 
\sum_{i,j}\bar{\Gamma}_{\beta,i}(p';E)
[G_{N^*}(E)]_{i,j}
\bar{\Gamma}_{\alpha,j}(p;E) 
\label{eq:tr}
\end{eqnarray}
with
\begin{eqnarray}
[G_{N^*}^{-1}]_{i,j}(E)&=& 
(E - m_{N^*_i})\delta_{i,j} - \Sigma_{i,j} (E)\,.
\label{eq:tr-g}
\end{eqnarray}
Here $i,j$ denote the bare $N^*$ states defined in the Hamiltonian.
$ m_{N^*_i}$ are their masses.
The first term (called meson-exchange amplitude from now on)
in Eq.(\ref{eq:fullt}) 
is defined by the following coupled-channels
equation
\begin{eqnarray}
t_{\beta,\alpha}(p',p;E)
& = & v_{\beta,\alpha}(p',p)
   +
\int_C dq q^2 \sum_{\gamma}
v_{\beta,\gamma}(p',q;E)G_\gamma(q,E)
t_{\gamma,\alpha}(q,p;E)
\label{eq:mext}
\end{eqnarray}
where $v_{\beta,\alpha}$ is defined by  meson-exchange mechanisms, and
$G_\gamma(q,E)$ is the propagator for channel $\gamma$.
The dressed vertexes and the energy shifts of the second term in 
Eqs.(\ref{eq:tr})-(\ref{eq:tr-g}) are defined by
\begin{eqnarray}
\bar{\Gamma}_{\alpha,j}(p;E) & = &
\Gamma_{\alpha,j}(p) + \int_C dq q^2 \sum_\gamma
t_{\alpha,\gamma}(p',q;E)G_\gamma(q,E)\Gamma_{\gamma,j}(q) 
\label{eq:dressf} \\
\Sigma(E)_{i,j} & = & \int_C dq q^2 \sum_\gamma
\Gamma_{\gamma,i}(q)G_\gamma(q,E)\bar{\Gamma}_{\gamma,j}(q),
\label{eq:selfe}
\end{eqnarray}
where $\Gamma_{\alpha,i}(p)$ defines the coupling of the
$i$-th bare $N^*$ state to channel $\alpha$.

Because of the quadratic relation between energy $E$ and momentum $p$,
there are two energy sheets for each two-body channel: the physical (unphysical) sheet
is identified with $Im(p) > (<)$ 0 for the stable two-particle
channels.  
 Thus the scattering
amplitudes of an $n$-channels model are defined on a Riemann energy
sheet which consists of $2^n$ sheets.
For the  EBAC-DCC model,
defined by Eqs.(\ref{eq:fullt})-(\ref{eq:selfe}),
each sheet can be defined
by symbol $(z_{\pi N}, z_{\eta N}, z_{\pi\pi N}, z_{\pi\Delta},
z_{\rho N}, z_{\sigma N})$ , where $z_\alpha $ could be
$p$ or $u$ representing the physical or unphysical sheets of
channel $\alpha$.
Note that an acceptable reaction model
can only have bound state poles and unitarity cuts on the physical sheet
$(pppppp)$.
The sheets from all other possible combinations of
$u$ and $p$  are called unphysical sheets on which
the scattering amplitude can have poles. We are however only interested
in poles which have large effects on scattering observables and therefore
they must be on the sheets which are near the  $(pppppp)$ physical sheet.
These poles are called resonance poles, and other poles are called shadow
poles. It is known\cite{taylor,mp87} that a shadow pole near the threshold of
a channel can also have large effects on scattering observables
and must also be considered in the search. 
As analyzed in
Ref.\cite{ssl09}  using several exactly soluble models, 
these poles are  in most cases on sheets where
the open(above threshold) meson-baryon channels are on
unphysical $u$ sheets and
the closed(below threshold) channels are on physical sheet.
In below we first recall how the  analytic continuation method we had
 developed in Ref.\cite{ssl09} is used to search for such
 resonance poles within
 EBAC-DCC model. 
We then describe how it is used to extract the residues of the
extracted resonance poles
from on-shell amplitudes.

\begin{figure}
\begin{center}
\includegraphics[width=8cm]{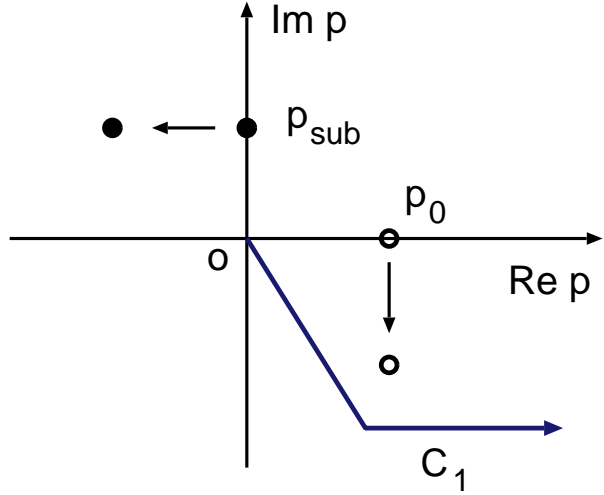}
\caption{The shift of the on-shell momentum (open circle/solid circle) of the 
two-particle Green function Eq. (\ref{eq-g-st}) as energy E
moves from a real value above/below the threshold energy 
to a complex value with negative imaginary part.
 $C^\prime_1$  is  the integration path 
for calculating Eqs.(\ref{eq:mext})-(\ref{eq:selfe})
amplitude for E on the unphysical  Riemann sheet.}
\label{fig:path}
\end{center}
\end{figure}

Since $v_{\alpha,\beta}$ and the bare vertex $\Gamma_{\alpha,i}$
are energy independent within the EBAC-DCC model , 
the analytic structure of the scattering
amplitude defined above as a function of $E$ is mainly determined by the
Green functions  $G_\gamma(q,E)$.
Thus the key for selecting the amplitude on physical sheet or
unphysical sheet
is to take an appropriate path of momentum integration $C$ in
Eqs.(\ref{eq:fullt})-(\ref{eq:selfe}) according to
the locations of the singularities of the meson-baryon
Green functions $G_{\alpha}(p,E)$ as $E$ moves to complex plane.
This can be done independently for each meson-baryon channel.
For  channel with stable particles such as $\pi N$ and $\eta N$,
the meson-baryon Green function is 
\begin{eqnarray}
G_{MB}(E,p) = \frac{1}{E - E_M(p) - E_B(p)} \,, \label{eq-g-st}
\end{eqnarray}
which has a pole at the on-shell momentum $p_0$ defined by
\begin{eqnarray}
E=\sqrt{m_M^2+p^2_{0}} + \sqrt{m_B^2+p^2_{0}}.
\label{eq-on-sh}
\end{eqnarray}
As an example, let us consider the analytic continuation
of the amplitude to the unphysical sheet of the $MB$ channel
when the energy $E$ is above the threshold  $Re(E)> m_B + m_M$
and $Im(E)<0$.
The on-shell momentum $p_0$ for such a $E$
is on the second and the fourth quadrant
of the complex momentum plane.
As $Im(E)$ becomes more negative 
as illustrated in Fig. \ref{fig:path}, the on-shell momentum (open
circle) moves into the fourth quadrant.
The amplitude on the unphysical sheet can be obtained by deforming
the path $C$ into  $C_1$
so that the on-shell momentum does not cross the integration contour.
For energy below the threshold for the MB channel ($E< m_B+m_M$),
the on-shell momentum $p_{sub}$ is on
the axis of positive imaginary. As the energy moves into the region 
of $Re(E)< m_B + m_M$ and $Im(E)< 0$, $p_{sub}$ moves
 to the second quadrant of complex p-plane and does not cross
path $C_1$,  as indicated by the dotted curves
in Fig.\ref{fig:path}.
Hence the amplitudes on the physical sheet of $MB$ channel
for energy below $MB$ threshold can also be obtained by taking the
path $C_1$.

\begin{figure}
\begin{center}
\includegraphics[width=6cm]{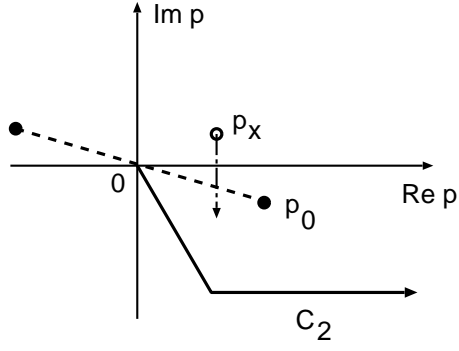}
\caption{ Contour $C_2$ 
for calculating Eqs.(\ref{eq:mext})-(\ref{eq:selfe})
for E on the unphysical Riemann sheet with the unstable particle
propagators, such as Eq.(\ref{eq-g-unst}) for $\pi \Delta$ channel.
See the text for the explanations of the dashed line and
the singularity $p_x$.}
\label{fig:c2}
\end{center}
\end{figure}

For the channels with unstable particle
 such as the $\pi \Delta$, as an example,
the Green function is of the following form
\begin{eqnarray}
G_{\pi\Delta}(E,p)  =  
\frac{1}{E- E_\pi(p) - E_\Delta(p)- \Sigma_{\Delta}(E,p)},
\label{eq-g-unst}
\end{eqnarray}
where
\begin{eqnarray}
\Sigma_\Delta(p,E)= \int_{C_3}  \frac{\{\Gamma_{\Delta,\pi N}(q)\}^2 q^2 dq}
{E-E_\pi(p)- [(E_\pi(q)+E_N(q))^2+p^2]^{1/2}}. \nonumber \\
\label{eq-sigma-pid}
\end{eqnarray}
The  $\pi\Delta$ Green function Eq. (\ref{eq-g-unst})
has a singularity at momentum $p=p_x$, which satisfies
\begin{eqnarray}
E-E_\pi(p_x)-E_\Delta(p_x)-\Sigma_\Delta(p_x,E)=0.
\label{eq-on-sh-pid}
\end{eqnarray}
Physically, this singularity corresponds to the $\pi\Delta$ two-body
'scattering state'.
There is also a discontinuity of the
$\pi\Delta$ Green function  associated with the $\pi\pi N$ cut
in $\Sigma_\Delta$, as shown in the dashed line in Fig. \ref{fig:c2},
where $p_0$ is defined by  
\begin{eqnarray}
E=E_\pi(p_0)+[(m_\pi+m_N)^2+p^2_0]^{1/2}.
\end{eqnarray}
Therefore, for $Re(E)> m_B + m_M, 2m_\pi + m_N$,
the integration contour $C$ 
must be chosen to be below the $\pi\pi N$ cut (dashed line)
 and the singularity $p_x$, such
as the contour $C_2$ shown in  Fig.\ref{fig:c2},
for calculating amplitudes on the unphysical sheet.

\begin{figure}
\begin{center}
\includegraphics[width=6cm]{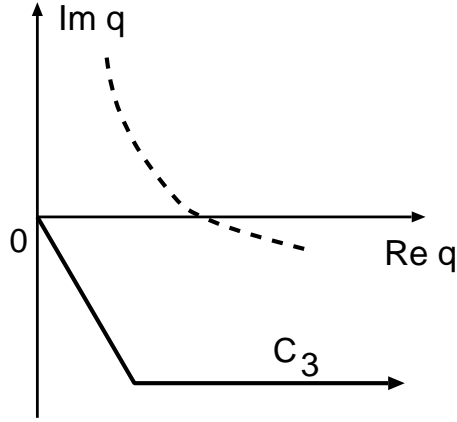}
\caption{ Contour $C_3$ for calculating the $\Delta$ self energy 
Eq.(\ref{eq-sigma-pid}) 
on the unphysical Riemann sheet. Dashed curve is the singularity $q_0$ of
the propagator in Eq. (\ref{eq:sing-q0}), which depends on
the spectator momentum $p$ on the contour $C_2$ of Fig.\ref{fig:c2}. }
\label{fig:c3}
\end{center}
\end{figure}

The singularity $q_0$ of the integrand of
Eq. (\ref{eq-sigma-pid})  depends on the spectator momentum $p$
\begin{eqnarray}
E-E_\pi(p) = [(E_\pi(q_0)+E_N(q_0))^2+p^2]^{1/2} \,.
\label{eq:sing-q0}
\end{eqnarray}
Thus  $q_0$ moves along the dashed curve, illustrated in Fig.\ref{fig:c3},
when the momentum $p$ varies along the path $C_2$ of Fig.\ref{fig:c2}.
To analytically continue $\Sigma_\Delta(p,E)$ to the unphysical sheet,
the contour $C_3$ of Eq. (\ref{eq-sigma-pid}) must be
 below $q_0$. A possible contour $C_3$ is the solid curve in Fig.\ref{fig:c3}.

\begin{figure}
\begin{center}
\includegraphics[width=10cm]{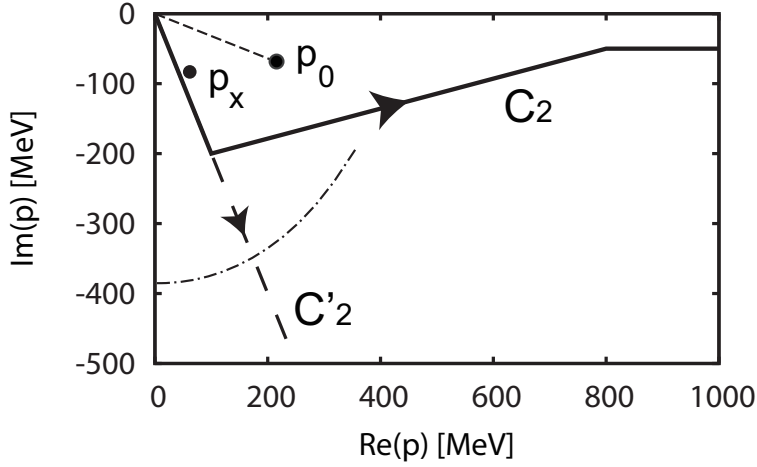}
\caption{The contour (solid curve) for calculating electromagnetic
matrix element. $p_0$ and $p_x$ are the singularities shown in 
Fig.\ref{fig:c2}. The dashed-dot curve is the singularity of the pion-exchange
$\gamma N \rightarrow \pi  \Delta$ matrix element at $E = (1357,  - 76 i) $ MeV.
\label{fig:actual-path} }
\end{center}
\end{figure}

\begin{figure}
\begin{center}
\includegraphics[width=6cm]{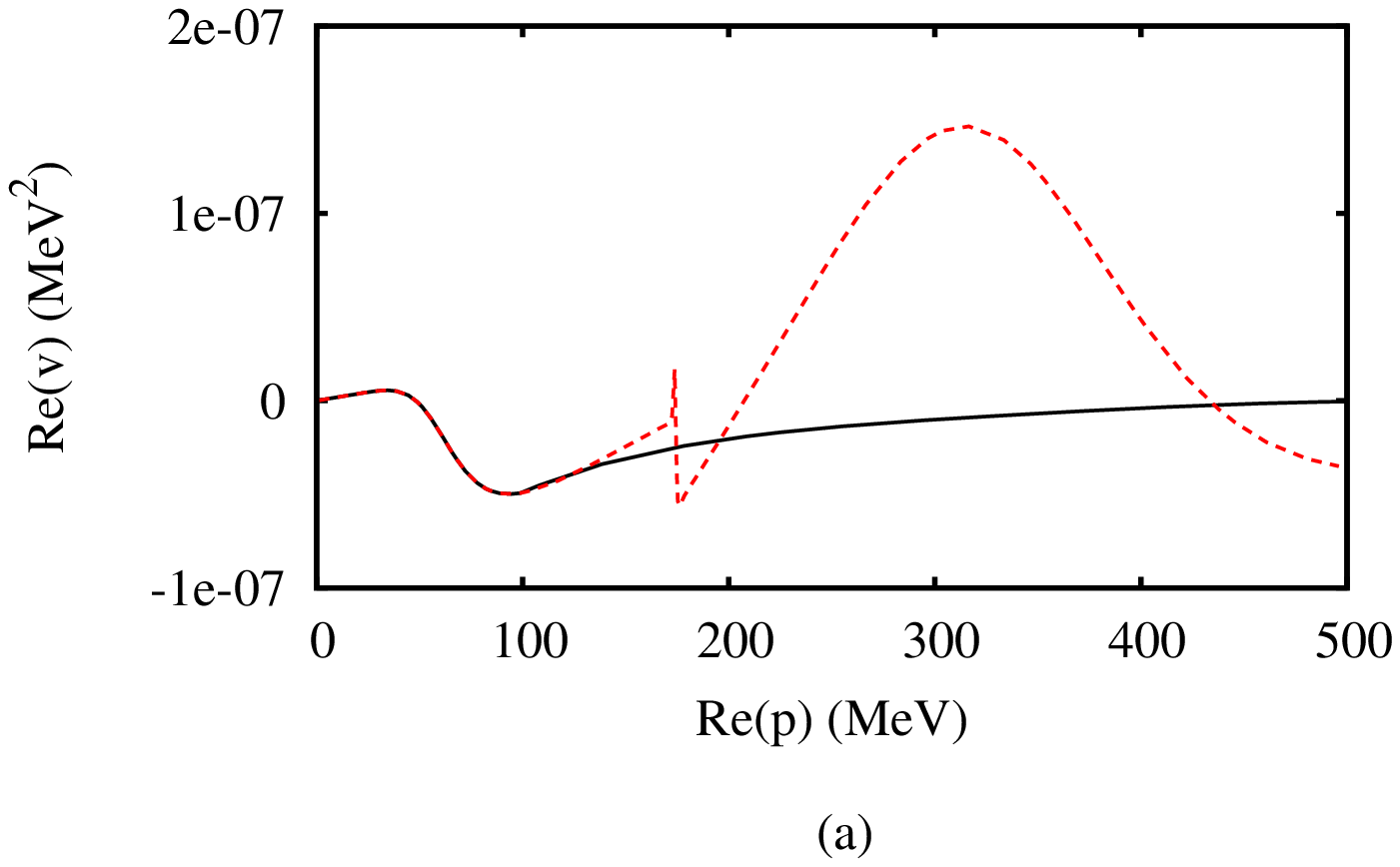}
\includegraphics[width=6cm]{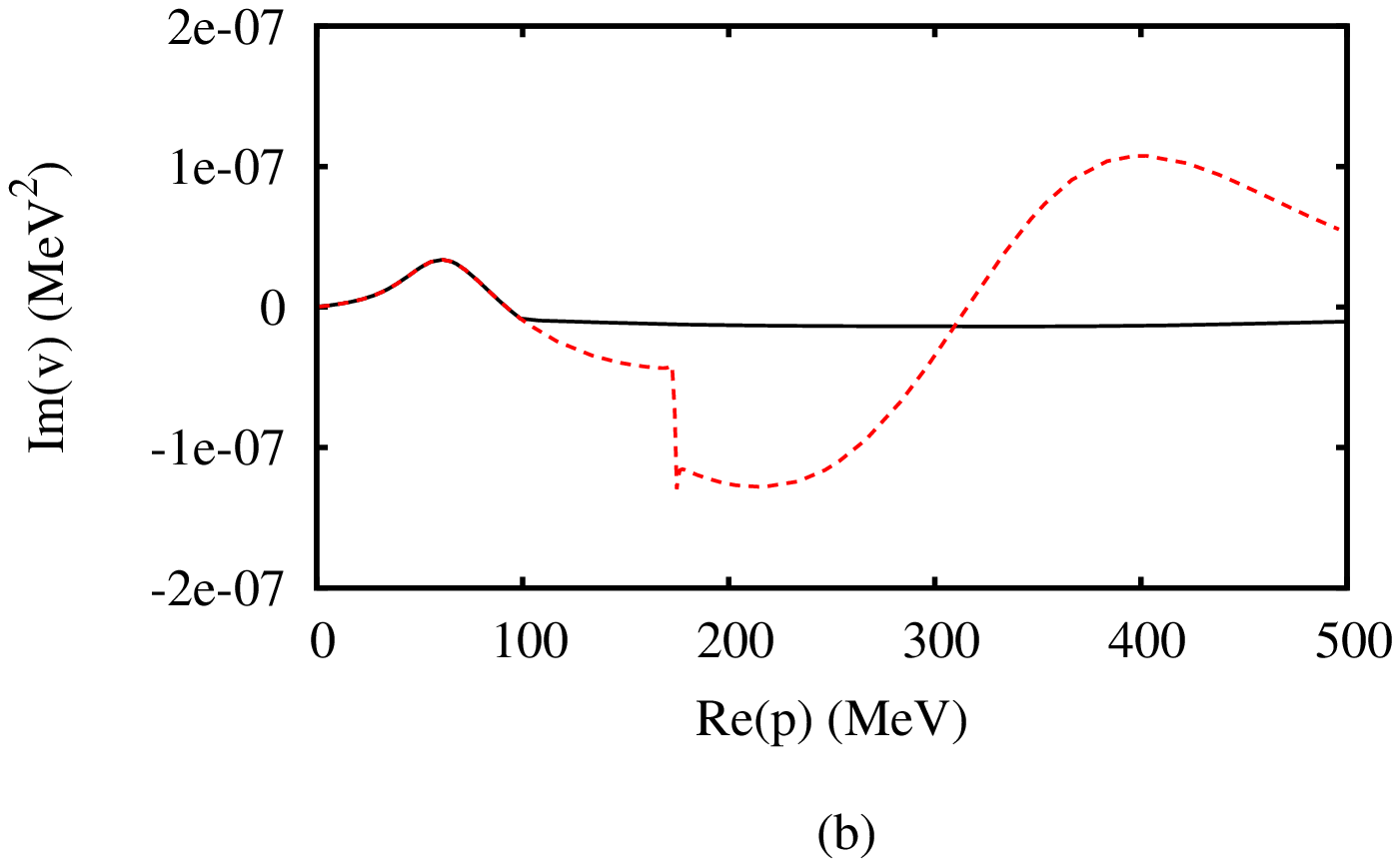}
\caption{The real (left) and the imaginary(right) parts
of  the half-off-shell matrix-elements of the
non-resonant potential $v_{\gamma N \rightarrow \pi\Delta}$ for
$P_{11}$ partial wave at $E = (1357, -76 i)$,
 as functions of the real part ($Re(p)$)
 of off-shell momentum.
The solid(dashed) curve is from  the calculation  
along the path $C_2$ ($C_2'$) shown in Fig.\ref{fig:actual-path}. 
\label{fig:gamnpid-path} }
\end{center}
\end{figure}

We emphasize here that we can deform the contour $C$
only in the region where the potential $v_{\alpha,\beta}(p',p)$
and the bare $N^*$ vertex $\Gamma_{MB,N^*}(p)$ 
are analytic. 
The contours described
above are chosen only from considering the singularities of $MB$ and $\pi\pi N$
Green functions. Thus they must be further
modified according to the analytic structure
of the considered $v_{\alpha,\beta}(p',p)$ and $\Gamma_{MB,N^*}(p)$
to obtain the scattering  amplitude
in the momentum region of interest.
This consideration is specially necessary when we need to get the
on-shell amplitude for extracting the residues of the identified resonance
poles.
The residue of the amplitude at resonance pole is
evaluated from the 'on-shell' matrix element, where the on-shell momenta
are defined as
$M_R = E_\pi(p^{on}_{\pi N}) + E_N(p^{on}_{\pi N})$ for $\pi N$ channel
and $M_R = q^{0,on}_{\gamma N} + E_N(q^{on}_{\gamma N})$ with
$Q^2= (q^{on}_{\gamma N})^2 - (q^{0,on}_{\gamma N})^2$ for $\gamma^* N$ channel.
Since  on-shell momentum are in general closer to the real axis than 
momentum on contour $C$, the analytic properties
of the meson-exchange potential has to be examined.
For example,
the t-channel meson exchange  potential
$v^t_{M'B',MB}(\vec{p}^{\,\,'},\vec{p})$ of the EBAC-DCC model
has  singularities at
\begin{eqnarray}
\Delta^2 - (\vec{p} - \vec{p}')^2 = 0
\end{eqnarray}
with $\Delta=E_{M'}(p')- E_M(p)$ or $E_{B'}(p')-E_B(p)$.
The form of $\Gamma_{MB,N^*}(p)$ is chosen such that its singularity is
at the pure imaginary momentum.
Thus the contours have to be chosen to also avoid these singularities.
As an example we show in Fig.\ref{fig:actual-path} 
the singularities associated with the $\pi\Delta$ channel at 
$E=(1357, -76i)$ MeV.
The dotted line for $\pi\pi N$ cut
and the circle shows $p_X$ are the singularities from the Green's function,
as discussed above.
The most relevant singularity of the meson-exchange
potential in our investigation of electromagnetic
pion production amplitude is due to
the t-channel pion exchange of $\gamma N \rightarrow
\pi\Delta$,
which is shown as the dashed-dot curve. Thus the
integration contour has to be modified to the solid curve $C_2$ in  
Fig.\ref{fig:actual-path}.
This can be understood from 
Fig. \ref{fig:gamnpid-path} in which we see that 
 the matrix element (dashed curves)
 of non-resonant potential
$v_{\pi\Delta,\gamma N}$ encounters the cut around $Re(p) \sim 170MeV$
with the path $C_2'$, but it varies smoothly (solid curves)
 along the path $C_2$.

\section{Extraction of Transition Form Factors}

 To indicate the essential features of our approach more clearly,
it is useful to first briefly describe how the resonance parameters are 
defined in the previous investigations.
The scattering amplitude $F_{\beta,\alpha}$ between any two channels
$\alpha$ and $\beta$ is related to the $S$-matrix element by
$F_{\beta,\alpha}= (1 - S_{\beta,\alpha})/2i$.
Within the rather general
theoretical framework discussed by, for example,
 Dalitz and Moorhouse\cite{dalitz},
Taylor\cite{taylorbook}, and McVoy\cite{mcvoy}, $F_{\beta,\alpha}$ at energies
 near a resonance pole position $M_R$
is parametrized 
as a sum of  a pole term and
a constant non-resonant contribution
\begin{eqnarray}
F_{\beta,\alpha}(E\rightarrow M_R) \sim \frac{R_{\beta,\alpha}}{M_R - E} +
 B_{\beta,\alpha},
\label{eq:f-pole}
\end{eqnarray}
where  $R_{\beta,\alpha}$ is the residue  at
the pole position $M_R$,  and the non-resonant amplitude $B_{\beta,\alpha}$
is  an  energy independent complex number.
By the unitarity condition imposed on the full $S$ matrix
$F_{\alpha,\beta}(E)$ at $E\rightarrow M_R$, the non-resonant term $B_{\alpha,\beta}$ is written in terms of
 an non-resonant S-matrix
$S^{B}$, which is unitary by itself ($S^{B} S^{B\dagger}=1$)
\begin{eqnarray}
B_{\beta,\alpha} & = & \frac{1 - S^{B}_{\beta,\alpha}}{2i}.
\end{eqnarray}
Then the pole term of Eq.(\ref{eq:f-pole}) 
 is  defined by the partial width $\Gamma_\alpha$
and a phase $\phi_\alpha$ arising from the presence of
the non-resonant term $B_{\alpha,\beta}$ 
\begin{eqnarray}
\frac{R_{\beta,\alpha}}{E - E_R} & = &
\frac{e^{i\phi_\beta}\sqrt{\Gamma_\beta/2}e^{i\phi_\alpha}\sqrt{\Gamma_\alpha/2}}
{E - M_R} .
\label{eq:f-p1}
\end{eqnarray}
 It is important to note that 
for the $\gamma N \rightarrow \pi N$ amplitudes we are going to consider,
$e^{i\phi_{\gamma N}}\sqrt{\Gamma_{\gamma N}}$ is the electromagnetic
$\gamma N\rightarrow N^*$ form factor which clearly must be a 
complex number when the the non-resonant term $B_{\pi N,\gamma N}$ is present
and  $\phi_{\gamma N} \neq 0$.
We will see that our formula are consistent with these earlier investigations
and will yield complex $\gamma N \rightarrow N^*$ form factors. Our main
advance is to provide their interpretations in terms of dynamics defined
within the EBAC-DCC model.  

 Here we  mention that by introducing appropriate 
energy-dependence of $Im(M_R)$, $R_{\alpha\beta}$,
and $B_{\alpha,\beta}$, the expression Eq.(\ref{eq:f-pole})
is used in practice to fit the experimental data.
This is the origin of
 the commonly used Breit-Wigner parametrization
 of the amplitude in physical energy region.
In some recent
analysis\cite{gwu-vpi,maid,inna} based on  such
a Breit-Wigner parametrization, 
the extracted $\gamma^* N \rightarrow N^*$ form factors are
 reported as 
real numbers. 
 Clearly, this is rather different from what one can interpret from
the above formula
used in the earlier analysis\cite{dalitz,taylorbook,mcvoy}.

We now explain that within the EBAC-DCC model, it is straightforward to
 extract the resonance parameters $M_R$, $B_{\alpha,\beta}$ 
and $R_{\alpha,\beta}$ of Eq.(\ref{eq:f-pole})
 by performing a Laurent expansion of the T-matrix defined 
Eqs.(\ref{eq:fullt})-(\ref{eq:selfe}).
We need to find
poles of scattering amplitudes $T_{\alpha,\beta}$.
In principle the pole of the scattering amplitude can be
found in the meson-exchange amplitude $t$ and/or
resonance amplitude $t^R$ of Eq. (\ref{eq:fullt}).
However as pointed out in Ref. \cite{juelich}, a pole $M_x$ of
the meson-exchange amplitude $t$ does not survive 
as a pole of the full amplitude when we introduce
coupling with bare $N^*$ states, since there is an exact cancellation
between the pole contributions from $t$ and $t^R$ at $E=M_x$.
Furthermore the non-resonant
term at resonance pole $t(E=M_R)$ is finite.
Thus, the resonance
poles of EBAC-DCC, or any model with bare $N^*$ states,
can be found by only analyzing $t^R$  defined
 by Eq.(\ref{eq:tr}). 
Consequently,  we only need to explain how
the residues of resonance poles are extracted from the term $t^R$.

The pole positions $M_R$ of $t^R$ are found from the zeros of the
determinant of $N^*$ propagator defined by Eq.(\ref{eq:tr-g})
\begin{eqnarray}
\Delta(E=M_R)= det[G_{N^*}^{-1}(E=M_R)]&=& 0 \,.
\label{eq:dete}
\end{eqnarray}
The pole term of the $N^*$ Green function
can be expressed as
\begin{eqnarray}
(G_{N^*}(E))_{ij} & = & \frac{\chi_i \chi_j}{E - M_R} \,,
\label{eq:pole-res}
\end{eqnarray}
where $i,j$ denote the bare $N^*$ state in the free Hamiltonian and
$\chi_i$ represents $i$-th 'bare' resonance component of the dressed $N^*$
and satisfies
\begin{eqnarray}
\sum_j (G_{N^*}(M_R)^{-1})_{ij}\chi_j & = &
\sum_j [(M_R - m_{N^*_i})\delta_{ij} - \Sigma(M_R)_{ij}]\chi_j = 0.
\label{eq:nstar-wf}
\end{eqnarray}
If there is only one bare $N^*$ state, 
{ with $G^{-1}_{N^*}(E)=1/(E-m_{N^*}-\Sigma(E))$},  
it is easy to see that
\begin{eqnarray}
\chi & = & \frac{1}{\sqrt{1 - \Sigma'(M_R)}} \,,
\end{eqnarray}
where $\Sigma'(M_R) = [d\Sigma/dE]_{E=M_R}$.
If we have two bare $N^*$ states, Eq.(\ref{eq:nstar-wf}) leads to 
\begin{eqnarray}
\chi_1 & = & \sqrt{\frac{M_R - m_{N^*_2} - \Sigma_{22}(M_R)}{\Delta'(M_R)}}
\,, \\
\chi_2 & = & \frac{\Sigma_{12}(M_R)}{M_R - m_{N^*_2} 
- \Sigma_{22}(M_R)}\chi_1
\end{eqnarray}
where $\Delta'(M_R)=[d\Delta/dE]_{E=M_R}$  can be evaluated using
Eq.(\ref{eq:dete}).

Now it is straightforward to see how the residues $R_{\beta,\alpha}$ 
and non-resonant term $B_{\beta,\alpha}$ of Eq.(\ref{eq:f-pole}) 
can be extracted from the amplitude $T_{\beta,\alpha}$ defined by
Eq.(\ref{eq:fullt}).  
   First we note that at $E$ near the resonance pole $M_R$,
the full amplitude defined by Eq.(\ref{eq:fullt}) can be written as
\begin{eqnarray}
T_{\beta,\alpha}(p^{on}_\beta,p^{on}_\alpha; E\rightarrow M_R)
                    & =& t_{\beta,\alpha}(p^{on}_\beta,p^{on}_\alpha; M_R) +
            t^R_{\beta,\alpha}(p^{on}_\beta,p^{on}_\alpha; E \rightarrow M_R)
\label{eq:fullt-pole}
\end{eqnarray}
where $p^{on}_\alpha$ is the on-shell momentum
of channel $\alpha$; e.g.
$M_R=E_\pi (p^{on}_{\pi N} )+E_N(p^{on}_{\pi N})$  for the $\pi N$ channel,
and $t_{\beta,\alpha}(p^{on},p^{on} ; M_R)$ is finite, as explained 
above.
By using Eq.(\ref{eq:tr}) for the definition of $t^R_{\beta,\alpha}$ and 
Eq.(\ref{eq:pole-res}) for the pole
term of $N^*$ propagator, we can perform Laurent expansion 
of the on-shell element of Eq.(\ref{eq:fullt-pole}) to obtain
\begin{eqnarray}
T_{\beta,\alpha}(p^{on}_\beta,p^{on}_\alpha ; E \rightarrow M_R)
= \frac{\bar{\Gamma}^R_\beta \bar{\Gamma}^R_\alpha}{E - M_R}
 + B_{\beta,\alpha} + B^1_{\beta,\alpha}(E - M_R) + ... \,.
\label{eq:fullt-exp}
\end{eqnarray}
where
\begin{eqnarray}
\bar{\Gamma}^R_\alpha & = & \sum_j \chi_j 
\bar{\Gamma}_{\alpha,j}(p^{on}_\alpha,M_R).
\label{eq:gamma-r}
\end{eqnarray}
Here the dressed vertex $\bar{\Gamma}_{\alpha,j}$ is defined by
Eq.(\ref{eq:dressf}).
The terms $B_{\beta,\alpha}$ and $B^1_{\beta,\alpha}$ in 
Eq.(\ref{eq:fullt-exp})  depend on the matrix elements of
meson-exchange amplitude $t$ of Eq.(\ref{eq:fullt})
\begin{eqnarray}
B_{\beta,\alpha} & = &
 t_{\beta,\alpha}(p^{on}_\beta,p^{on}_\alpha ; M_R)
 + \frac{d}{dE}\left[(E - M_R)
 t^R_{\beta,\alpha}(p^{on}_{\beta},p^{on}_{\alpha};E)\right]_{E=M_R}.
\label{eq:c0}
\end{eqnarray}
The term $B^1_{\beta,\alpha}$ can be calculated, but is not relevant to our
following discussions. 

Let us now consider Eq.(\ref{eq:fullt-exp}) for $\alpha=\beta=\pi N$ case.
We  need to relate the residue $\bar{\Gamma}_{\pi N} \bar{\Gamma}_{\pi N}$
 of its pole term 
to the  residue of the $\pi N$ elastic scattering amplitude $F_{\pi N,\pi N}$  
defined by  the standard notation
\begin{eqnarray}
F_{\pi N,\pi N} (E) & = & \frac{S_{\pi N, \pi N} (E) - 1}{2i}  
 = [\frac{R e^{i \phi}}{M_R - E}]_{E\rightarrow M_R} \,,
\label{eq:def-reds}
\end{eqnarray}
where $S_{\pi N,\pi N}$ is the partial-wave S-matrix. In terms of the 
normalization of EBAC-DCC model 
$S_{\pi N,\pi N}(E) = 1- 2 i[\pi p^{on}E_\pi(p^{on})E_N(p^{on})/E]
T_{\pi N,\pi N}(p^{on},p^{on} ; E)$ , we find that 
($p^{on}$ stands for $p^{on}_{\pi N}$)
\begin{eqnarray}
F_{\pi N,\pi N} (M_R) = -
\pi \frac{p^{on} E_N(p^{on})E_\pi(p^{on})}{M_R}T_{\pi N,\pi N}(p^{on},p^{on},M_R)
\end{eqnarray}
Keeping only the pole term of Eq.(\ref{eq:fullt-exp}) in evaluating
the above equation and using the definition Eq.(\ref{eq:def-reds}),
we then obtain
\begin{eqnarray}
R e^{i \phi} & = & 
\pi \frac{p^{on} E_N(p^{on})E_\pi(p^{on})}{M_R} \bar{\Gamma}^R_{\pi N}
\bar{\Gamma}^R_{\pi N} \,.
\label{eq:pin-resid}
\end{eqnarray}
The $\pi N$ elasticity  of a resonance is then defined as
\begin{eqnarray}
\eta_e= \frac{R}{-Im (M_R)}
\label{eq:elasticity}
\end{eqnarray} 

 With the similar procedure, we  can perform the Laurent expansion of
$\gamma^* N \rightarrow \pi N$ amplitude to obtain
\begin{eqnarray}
T_{\pi N, \gamma N}(p^{on}, q^{on} ; E\rightarrow M_R) =
\frac{\bar{\Gamma}^R_{\pi N}(p^{on}) \bar{\Gamma}^R_{\gamma N}(q^{on}, Q^2)}
{E-M_R} +
B_{\pi N, \gamma N} + \cdot\cdot\cdot
\label{eq:em-exp-1}
\end{eqnarray}
where $q^{on}$ is the $\gamma N$ on shell momentum defined by
$M_R=q_0+E_N(q^{on})$ and the momentum-transfer $Q^2=(q^{on})^2 - q^2_0$.
As discussed in section I, a nucleon resonance
 can be interpreted\cite{bohm,dalitz} as 
an "eigenstate " of  the Hamiltonian $H|\psi^R_{N^*}> = M_R |\psi^R_{N^*}>$. 
Then 
from the spectral expansion of the Low Equation for reaction amplitude 
$T(E)= H' + H'\frac{1}{E-H}H'$, where we have defined $H'=H-H_0$ with
  $H_0$ being the non-interacting free
Hamiltonian, we have
\begin{eqnarray}
T_{\pi N, \gamma N}(p^{on}, q^{on} ; E\rightarrow M_R)
&=&\frac{<p^{on}|H'|\psi^R_{N^*}><\psi^R_{N^*}|H'|q^{on},Q^2>}{E-M_R} + \cdot\cdot 
\label{eq:em-exp-2}
\end{eqnarray}
 Obviously, we can see that
$ <\psi^R_{N^*}|H'|q^{on},Q^2> = <\psi^R_{N^*}| J^\mu(Q^2)\epsilon_\mu |N>$ 
is determined by
the electromagnetic current operator $J^\mu(Q^2)$. It must be a complex
number since the resonance wavefunction $\psi^R_{N^*}$ contains scattering
states.
Comparing Eqs.(\ref{eq:em-exp-1}) and 
(\ref{eq:em-exp-2}), we then interpret
$\bar{\Gamma}^R_{\gamma N}(q^{on},Q^2)$ as 
the $N^*_R \rightarrow \gamma^* N$ transition form
factor. As seen in Eq.(\ref{eq:pole-res}), the resonance consists of
all bare $N^*$  components and hence
we have
\begin{eqnarray}
<\psi^R_{N^*}| J^\mu(Q^2)\epsilon_\mu |N> = \sum_{i}\chi_i 
\bar{\Gamma}_{\gamma N, N^*_i}(q^{on},Q^2) 
\label{eq:dress-ff}
\end{eqnarray}

Using the normalizations defined
in Ref.\cite{jklmss09} and following the definition originally
introduced for the constituent quark model\cite{copley}, 
the usual $\gamma^* N \rightarrow N^*$ transition form factors 
are related to our extracted from factors by
\begin{eqnarray}
A_{3/2}(Q^2) & = & C \sum_j \chi_j \bar{\Gamma}^R_{\gamma^*
 N,j}(Q^2,M_R,\lambda_\gamma=1,\lambda_N=-1/2) \, \label{eq:emff-a32},\\
A_{1/2}(Q^2) & = & C 
\sum_j \chi_j \bar{\Gamma}^R_{\gamma^*
 N,j}(Q^2,M_R,\lambda_\gamma=-1,\lambda_N=-1/2) \, \label{eq:em-a12},\\
S_{1/2}(Q^2) & = & C \sum_j \chi_j \bar{\Gamma}^R_{\gamma^*
 N,j}(Q^2,M_R,\lambda_\gamma=0,\lambda_N=-1/2) \, \label{eq:em-c12},\\
\end{eqnarray}
where  $\lambda_N$ and $\lambda_\gamma$ are the helicities of the initial
nucleon and photon, respectively, and
\begin{eqnarray}
C=\sqrt{\frac{E_N(\vec{q})}{m_N }} \frac{1}{\sqrt{2K}} \times
\sqrt{\frac{(2j+1)(2\pi)^3(2q_0)}{4\pi}}
\label{eq:coef-c}
\end{eqnarray}
where $K  =  (M_R^2 - m_N^2)/(2M_R)$.

\section{Results and Discussions}

 In this section, we illustrate our procedures by presenting
the results for the pronounced  resonances
in $P_{33}$, $D_{13}$ and the  complex $P_{11}$ partial waves. 
 We also investigate the extent to which our results can be compared
with those extracted from using Breit-Wigner form of resonant amplitudes
to fit the data.
 
 Before we present our results for electromagnetic form factors,
it is useful to first discuss our results from 
$\pi N$ scattering amplitudes, which were briefly presented in 
Ref.\cite{ssl09,sjklms10}.
The  extracted pole positions ($M_R)$ and elasticities $\eta_e$ defined by
Eq.(\ref{eq:elasticity}) for $P_{33}$, $D_{13}$ and $P_{11}$
are compared with the values from Particle Data Group\cite{pdg}
in Table \ref{tab:poles}. We see that our results correspond well
with PDG, while only one $P_{11}$ near 1360 MeV is listed by
Particle Data Group ( PDG) \cite{pdg}.
The extracted residues $R e^{i\phi}$, defined in Eq.(\ref{eq:pin-resid}),
 for $\pi N$ amplitude
 are compared with some of the previous works in Table
\ref{tab:residues-pin}. We see that the agreement in $P_{33}$ and $D_{13}$
are excellent.
 However, we see that the
residues of the $P_{11}$  resonances extracted by four groups
do not agree well while we agree well with
GWU/VPI only for the resonance at  1356 MeV.
\begin{table}[t]
\caption{The extracted resonance poles ($Re M_R, -Im M_R)$ MeV and
elasticity $\eta_e$ (Eq.(\ref{eq:elasticity})) are compared with the
values listed by PDG\cite{pdg}.
 }
\begin{tabular}{|c|c|c|c|c|c|}
\hline
  &   $M_R$ (EBAC-DCC)& location  &     $M_R$ (PDG) &  $\eta_e$ (EBAC-DCC)& $\eta_e$ (PDG) \\ \hline
$P_{33}$  & (1211, 50) & (u-ppp-)&  (1209 - 1211 , 49 - 51)& 100$\%$ &  100 $\%$ \\ \hline
$D_{13}$  & (1527, 58) &(uuuupp) & (1505 - 1515  , 52 - 60)&  65 $\%$ & 55 - 65 $\%$\\ \hline
$P_{11}$ &  (1357, 76)&(upuupp)  & (1350 - 1380, 80 - 110) & 49 $\%$ & 55 - 75  $\%$\\
         & (1364,106)  & (upuppp)                       &  &  60 $\%$ & \\
         & (1820, 248)& (uuuuup)  & (1670 - 1770, 40 - 190) &  8 $\%$ & 10 - 20 $\%$ \\
\hline
\end{tabular}
\label{tab:poles}
\end{table}

 \begin{table}[t]
\caption{The extracted $\pi N$ residues $R e^{i\phi}$ defined by 
Eq.(\ref{eq:pin-resid}) are compared with several previous results.}
 \begin{tabular}{|c|cc|cc|cc|cc|}
\hline
 & EBAC-DCC &        & GWU-VPI\cite{gwu-vpi}  &    & Cutkosky\cite{cut}& 
&J\"{u}lich\cite{juelich} & \\
 & R & $\phi$ & R & $\phi$ & R & $\phi$ &R & $\phi$  \\
 \hline
 $P_{33}(1210)$ & 52& -46  &  52&  -47    &  $53$ & $-47$& 47 & -37
  \\
 \hline
 $D_{13}(1521)$ &  38&  7  & 38 & -5    &  $35$  &$-12$ & 32 & -18 \\
 \hline
 $P_{11}(1356)$ & 37  & -111  &  38  & -98    &  $52$ & $-100$& 48 & -64\\
 $\,\,\,\,\,\,\,\,(1364)$ &   64  & -99  & 86 & -46    &  - & -& - & - \\
 $\,\,\,\,\,\,\,\,(1820)$ &   20 & -168  & - & -    & 9   & -167 &-&- \\
 \hline
 \end{tabular}
\label{tab:residues-pin}
 \end{table}

In Table \ref{tab:poles},
we also indicate the location of each pole on Riemann energy sheet.
Since we only search for
poles in the region where the open (above threshold)
channels are on unphysical $u$ sheet and close channels (below threshold)
on physical $p$ sheets,
as described in section II,
the quantity  deciding which sheet each resonance in Table \ref{tab:poles}
is on are
the branching points for each channel, Within JLMS fit they
are ($1077, 1486, 1216, 1363-33i, 1703-75i, 1906-323i $) MeV
 for
($\pi N, \eta N, \pi\pi N, \pi\Delta,
\rho N, \sigma N$ ) , respectively. 
For example, the $P_{11}$ pole at 1357 MeV ( 1364 MeV)
is below (above)  the $\pi\Delta$ threshold $1363$ MeV and is on
$upuuupp$ $(upuppp)$ sheets since both poles are above $\pi N$ and 
$\pi\pi N$ channels and below $\eta N, \rho N$ and $\sigma N$ channels.
Thus their residues 
are very different although their positions are very close,
since they are on different Riemann sheets.
These two-poles structure near the $\pi\Delta$ threshold are also
found in  the  earlier analysis of VPI\cite{vpi84} and
Cutkosky and Wang\cite{cut}, and the recent analysis by the
GWU/VPI\cite{gwu-vpi}
and J\"{u}lich\cite{juelich} groups.

Our results presented in Tables I and II suggest that the  
resonance parameters of the pronounced and well isolated resonance poles, 
such as $P_{33}(1210)$
and $D_{13}(1527)$, are rather safely determined by the structure of
the empirical partial wave amplitudes 
as far as the employed models
have  the correct analytic properties
in the region not far from the physical region.
On the other hand, the residues of poles near threshold
are sensitive to the dynamical content of the models, as we have seen
in the considered $P_{11}$ case.

\begin{figure}
\includegraphics[width=5cm]{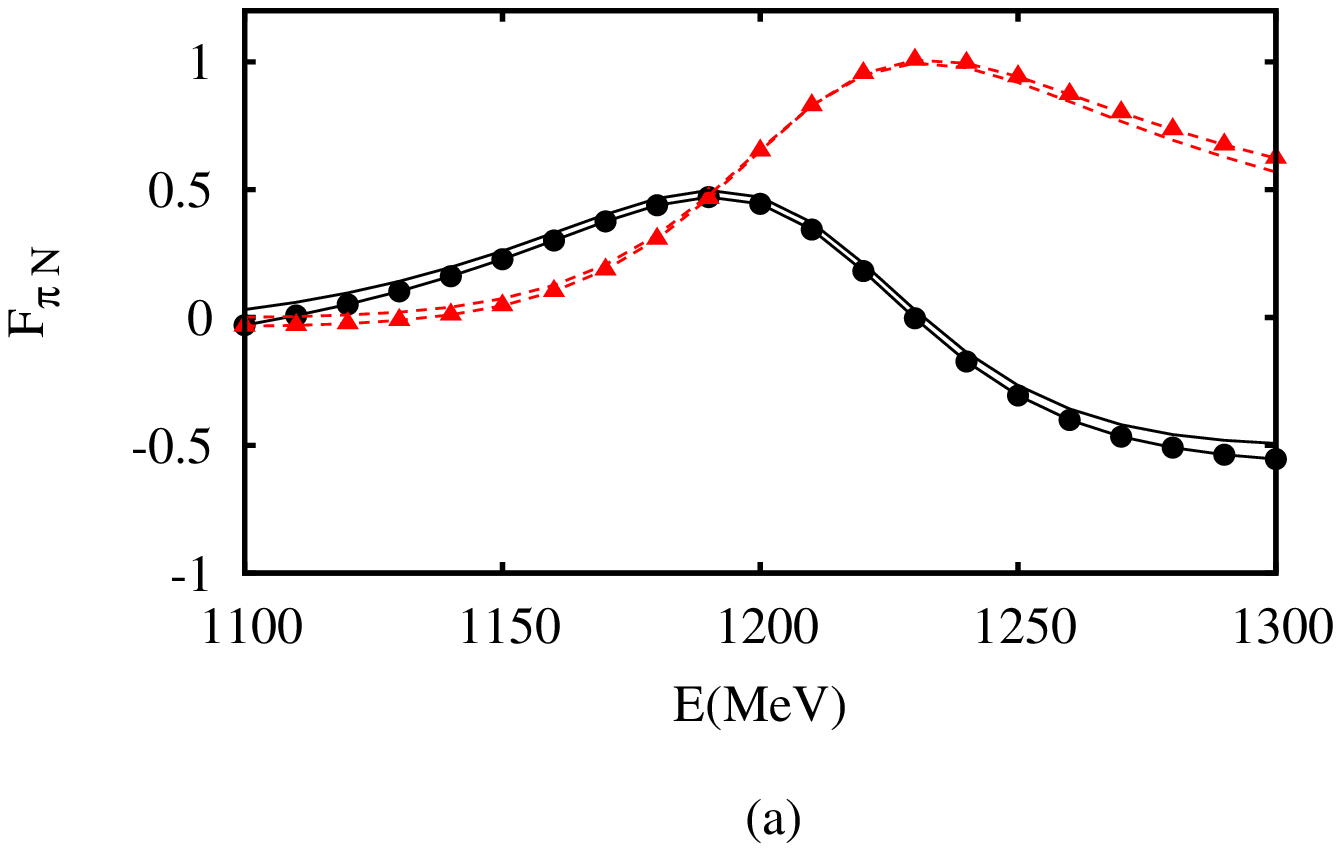} \hspace*{0.1cm}
\includegraphics[width=5cm]{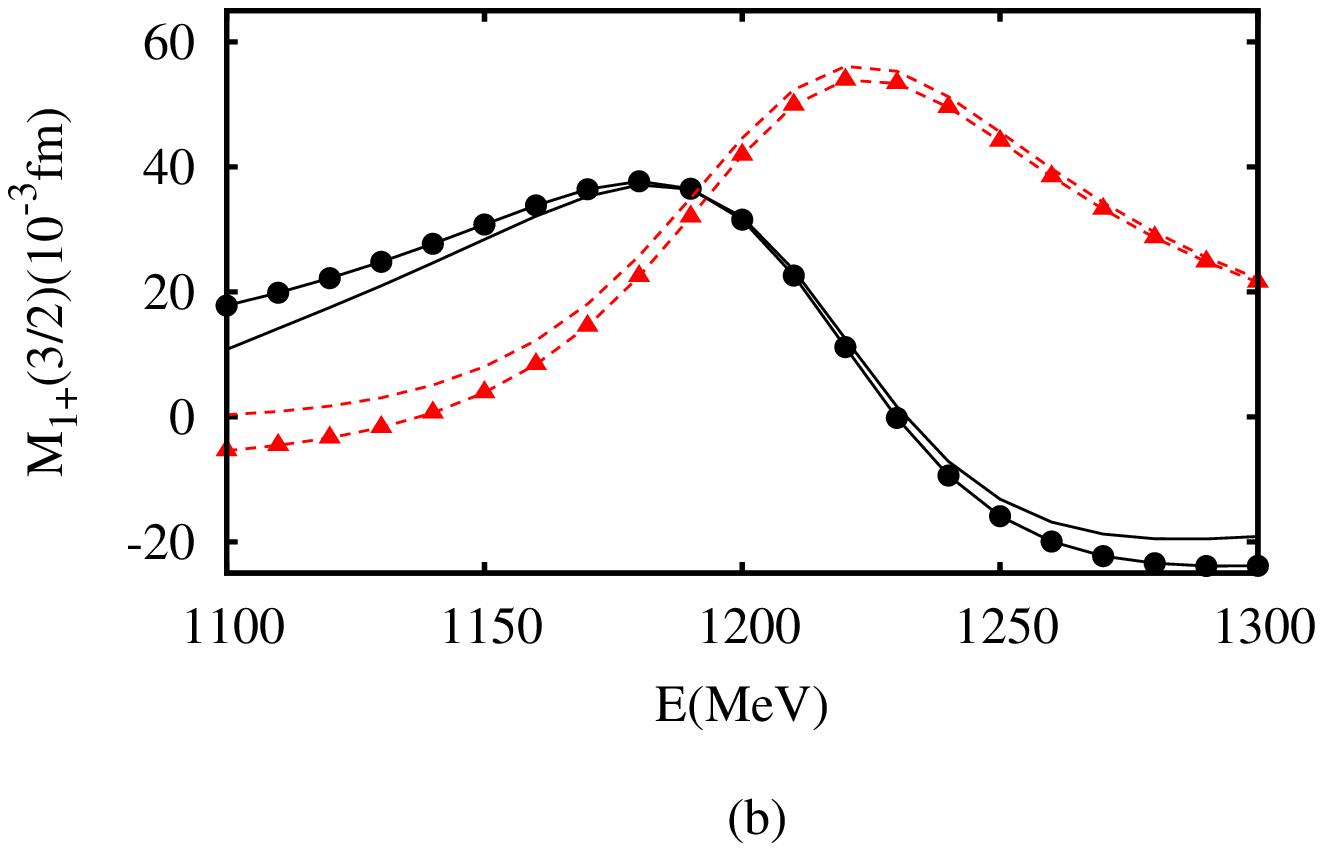} \hspace*{0.1cm}
\includegraphics[width=5cm]{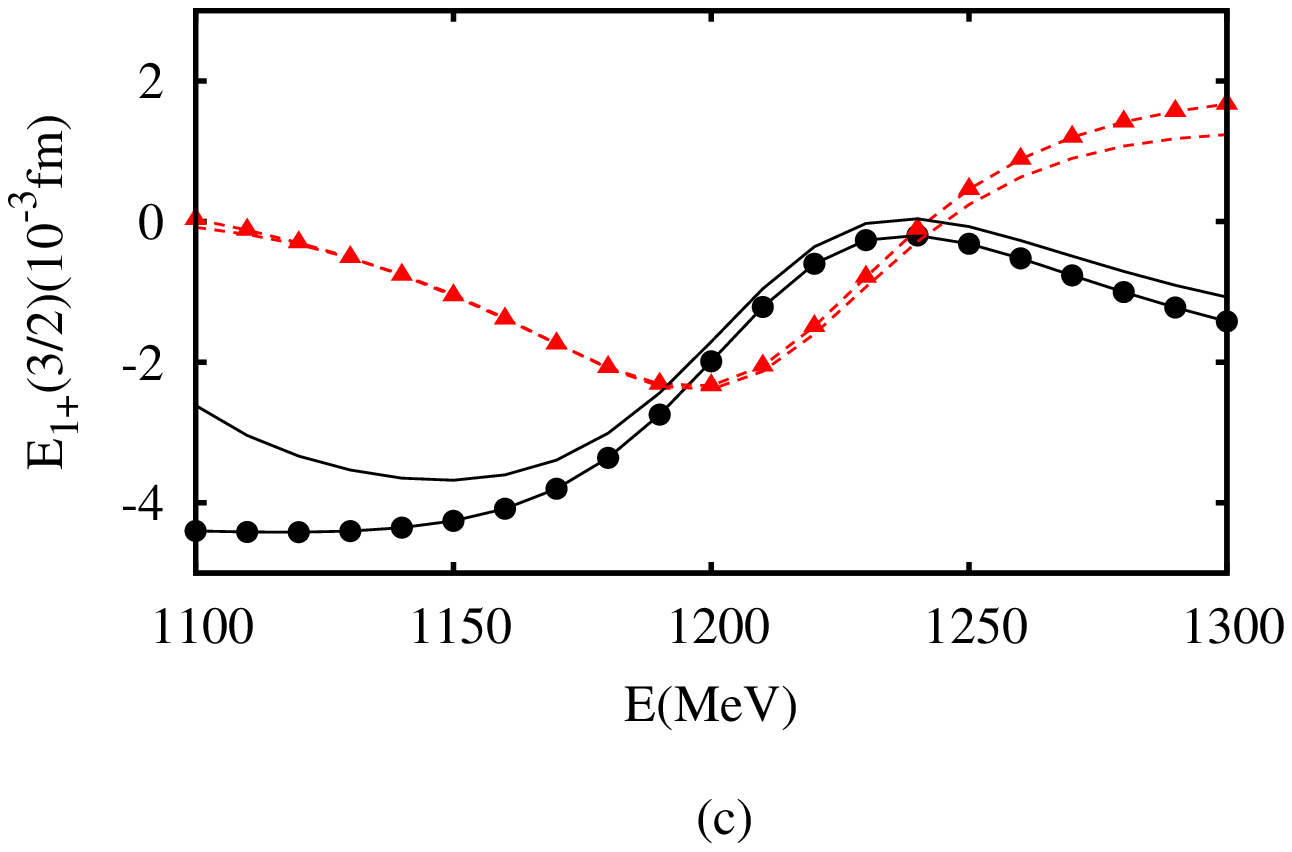}
\caption{Energy dependence of
the $\pi N$ amplitude(a)
and the $\gamma\pi$  $M_{1+}(3/2)$(b)  and $E_{1+}(3/2)$(c)   amplitudes
 of $P_{33}$ channel.
The solid circle(triangle) shows real(imaginary) part of the
amplitude calculated using Eq. (\ref{eq:f-pole}).
The solid (dashed) curve shows real(imaginary) part of the amplitude
of EBC-DCC model.}
\label{fig:p33-pole-approx}

\end{figure}

\begin{figure}
\includegraphics[width=5cm]{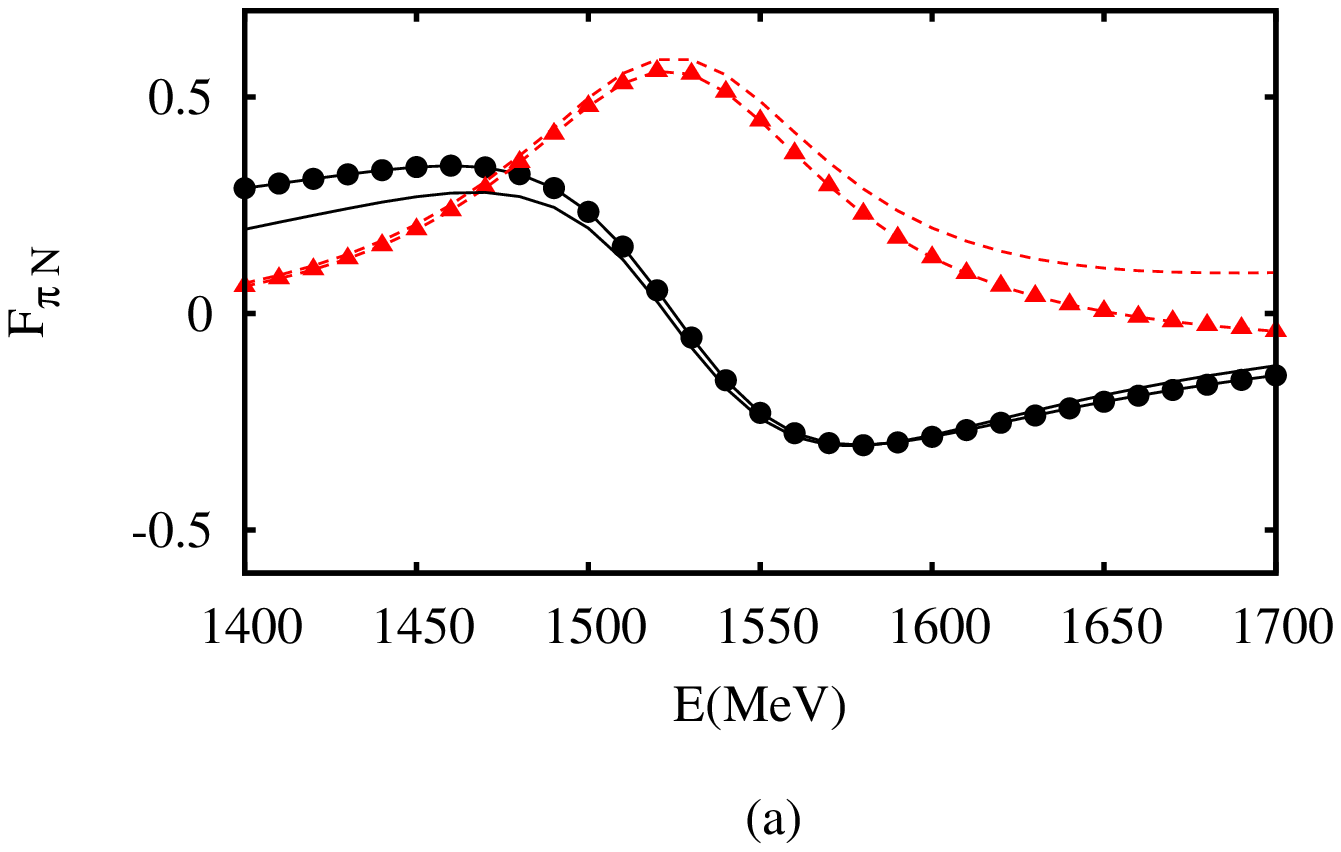} \hspace*{0.1cm}
\includegraphics[width=5cm]{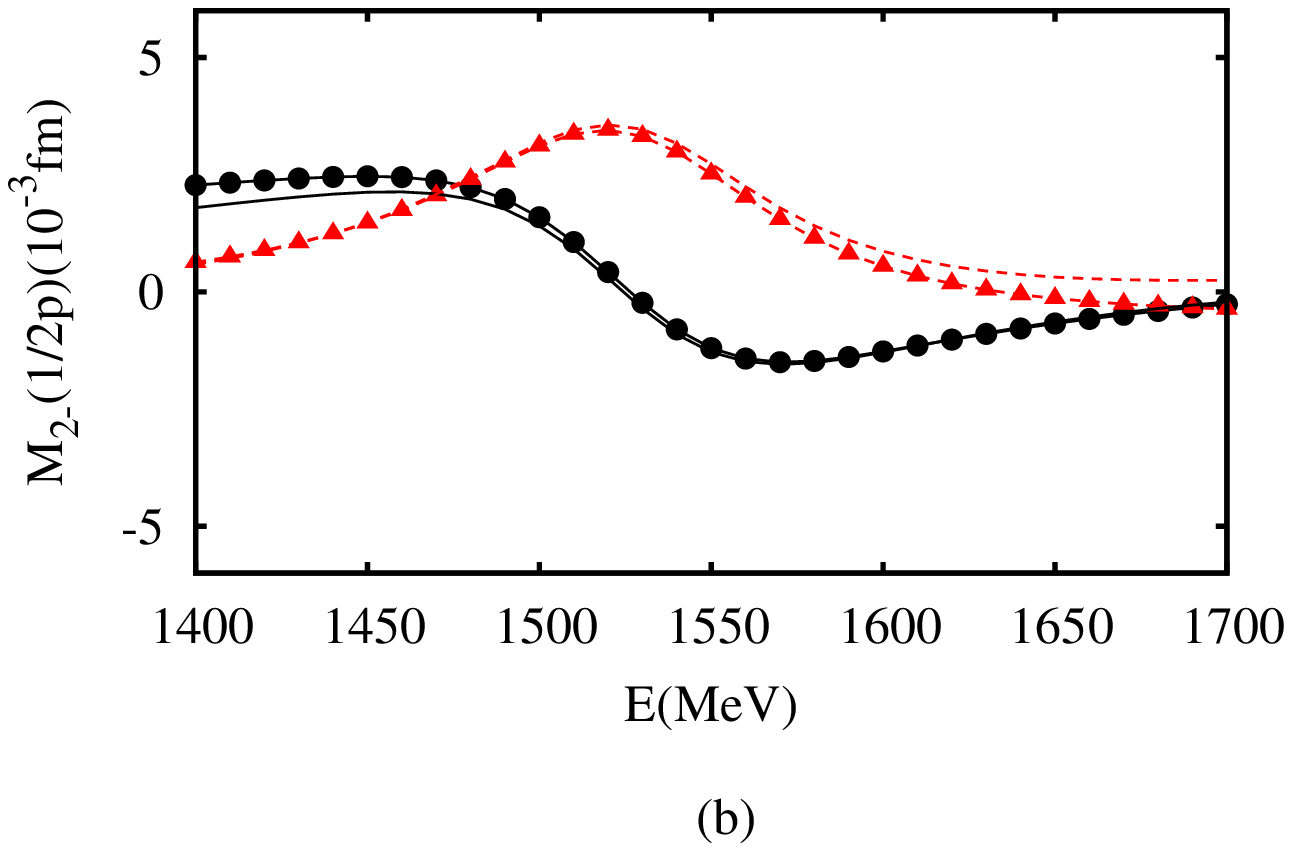} \hspace*{0.1cm}
\includegraphics[width=5cm]{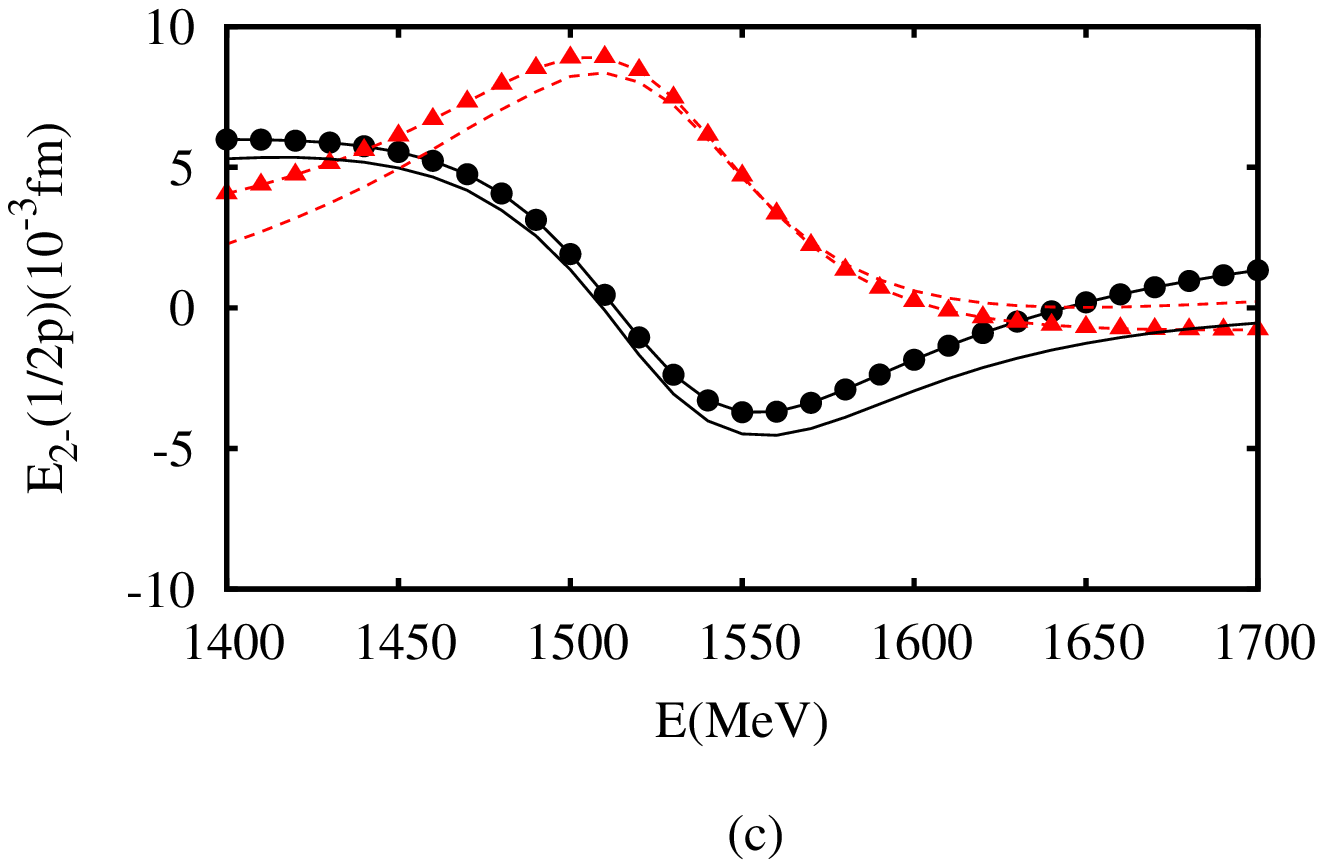}
\caption{Energy dependence of the
the $\pi N$ amplitude(a)
and the $\gamma\pi$  $M_{2-}(1/2p)$(b)  and $E_{2-}(1/2p)$(c)   amplitudes
 of $D_{13}$ channel.
The solid circle(triangle) shows real(imaginary) part of the
amplitude calculated using Eq. (\ref{eq:f-pole}).
The solid (dashed) curve shows real(imaginary) part of the amplitude
of EBC-DCC model.}
\label{fig:d13-pole-approx}
\end{figure}

 We now turn to presenting our results for $\gamma^* N \rightarrow N^*$
form factors $A_\lambda(Q^2)$ and $S_\lambda(Q^2)$. We first observe that
for the  isolated resonances in $P_{33}$ and $D_{13}$, Eq.(\ref{eq:fullt-exp}) 
and Eq.(\ref{eq:c0}) for $\gamma^* N \rightarrow \pi N$ multipole amplitudes at
 $E \rightarrow M_R$
can be  approximated as the following simple form
\begin{eqnarray}
T_{\pi N,\gamma N}( E\rightarrow M_R) =  B_{\pi N,\gamma N}
- \frac{R_{\pi N,\gamma N}}{E-M_R}\,,
\label{eq:res-pole}
\end{eqnarray}
where the complex constants are evaluated at resonance
position $E=M_R$ 
\begin{eqnarray}
B_{\pi N,\gamma N} & = &
 t_{\pi N,\gamma N}(p^{on},q^{on} ; M_R)
 + \frac{d}{dE}\left[(E - M_R)
 t^R_{\pi N,\gamma N}(p^{on},q^{on};E)\right]_{E=M_R}\,,\\
R_{\pi N,\gamma N}&=&\bar{\Gamma}^R_{\pi N}(p^{on},M_R)A_\lambda(Q^2,M_R)/C\,,
\end{eqnarray}
 where $C$ is defined by Eq.(\ref{eq:coef-c}).
We observe that the expression Eq.(\ref{eq:res-pole}), evaluated  with all  constants
 except $E$ kept at their complex values  at pole position $M_R$, is a good approximation in the
physical region of $E$ near $W_R=Re(M_R)$.  Similar good approximation
is also for the $\pi N \rightarrow \pi N$ amplitudes, 
 as also reported in Ref.\cite{juelich}. Our findings 
are shown 
in Figs.\ref{fig:p33-pole-approx} and \ref{fig:d13-pole-approx} for 
the $P_{33}$ and $D_{13}$ partial waves, respectively. The determined
constants $B_{\pi N,\gamma^*N}$, 
$R_{\pi N,\gamma N}$, $B_{\pi N,\pi N}$ $R_{\pi N,\pi N}$, and $M_R$
for each case in  
Figs.\ref{fig:p33-pole-approx} and \ref{fig:d13-pole-approx}
are presented in Table  \ref{tab:laurent}.

\begin{table}[t]
\caption{Extracted resonance parameters.  $R_{\beta,\alpha}$ and
$B_{\beta,\alpha}$ are for the $\pi N $ elastic scattering amplitude
 $F_{\pi N,\pi N}$ 
and multipole amplitudes $E_{L \pm }, M_{L \pm }$ of the pion photoproduction.}
\begin{tabular}{|c|c|cc|ccc|}\hline
          & $M_R(MeV)$ & $R_{\pi N,\pi N}(MeV)$ & $B_{\pi N,\pi N}$ 
           &  & $R_{\pi N,\gamma N}(10^{-3}fm MeV)$ & $B_{\pi N,\gamma N}(10^{-3}fm)$ \\ \hline
$P_{33}$  & 1211 - 50i     &   36.1 - 37.7i  &  -0.43  + 0.13i  
          & $M_{1+}(3/2)$ &  -2728 + 1436i    &   -7.43 - 3.86i\\ 
           &               &                 &
          & $E_{1+}(3/2)$ &  175 + 118i  &   -3.49 + 1.51i\\ \hline
$D_{13}$  & 1527 - 58i     &  37.6 + 4.9i    & 0.06 - 0.08i
         & $M_{2-}(1/2p)$ &  -224 - 61.6i & 1.01 - 0.44i\\
         &                 &                 &
         & $E_{2-}(1/2p)$ &  -437 - 368i & 4.25 + 0.36i \\ \hline
\end{tabular}
\label{tab:laurent}
\end{table}


We now note that the expression Eq.(\ref{eq:res-pole}) looks similar to the
commonly used amplitude with a
Breit-Wigner parametrization
\begin{eqnarray}
T^{BW}_{\pi N, \gamma N}(E) & = & B_{\pi N, \gamma^*N}(Q^2 ; E) + 
\frac{\Gamma^{1/2}_{\pi N}(E)e^{i\phi^{BW} (E)} A^{BW}_\lambda(Q^2,E)}
{E - (W_R - i\frac{ \Gamma_{tot}(E)}{2})}.
\label{eq:res-bw}
\end{eqnarray}
where $\Gamma_{tot}(E=W_R)$ and $\Gamma_{\pi N}(E=W_R)$
are called the total width and partial decay width for $\pi N$ channel, 
respectively, and $A^{BW}_\lambda(Q^2,q,E)$ is assumed to be real numbers.
 The energy dependence of these widths as well as the phase 
factor $\phi^{BW} (E)$
are parts of the assumptions in those analysis, which of course will influence
how the non-resonant amplitude $B_{\pi N,\gamma^*N}(Q^2 ; E)$ is adjusted
to fit the data.

Eqs.(\ref{eq:res-pole}) and (\ref{eq:res-bw}) have similar structure, but they
have important differences. First  Eq.(\ref{eq:res-pole}) is evaluated
at complex $M_R$ and hence the on-shell momentum $q^{on}$ and $p^{on}$
are also complex.
On the other hand, all energy and momentum variables 
in Eq.(\ref{eq:res-bw}) are real numbers defined by the physical
energy $E$. The non-resonant amplitude $B_{\pi N,\gamma N}$
 in Eq.(\ref{eq:res-pole}) is obtained from a coupled-channel
calculation, while $B_{\pi N,\gamma^*N}(Q^2 ; E)$ in Eq.(\ref{eq:res-bw})
is often calculated from tree-diagrams
of phenomenological Lagrangian with unitarization using $\pi N$
amplitude.  
Thus it is difficult to see that the helicity amplitudes of 
$\gamma^* N \rightarrow N^*$ extracted from these
two rather different approaches can be compared.  

The two-pole structure of $P_{11}$ resonances near $\pi\Delta$ threshold
poses a problem
in interpreting our results for the
$\gamma^* N \rightarrow N^*$ form factors $A_\lambda(Q^2)$.
We note that  
Eqs.(\ref{eq:res-pole}) is valid for each of these two poles, but
they are on different Riemann surfaces.  
Thus we need to find a parametrization which carries the sheet information in
representing these two-pole contributions.
Here we follow the approach of Refs. \cite{kato65,fujii1,fujii2}
and  a similar formula used in extracting meson
resonances\cite{mp87,bugg}.

 We first use Eq.(\ref{eq:res-pole}) to write
the $\pi N \rightarrow \pi N$ and $\gamma N \rightarrow \pi N$
scattering amplitudes on the $\pi\Delta$ physical($a=p$)
and unphysical($a$=$u$) sheet as
\begin{eqnarray}
T^{(a)}_{\beta,\alpha}(p^{on}_\beta,p^{on}_\alpha, E \rightarrow M^{(a)}_R)
 & = & - \frac{R^{(a)}_{\beta,\alpha}}
{E - M^{(a)}_R} + B^{(a)}_{\beta,\alpha},
\end{eqnarray}
where $\alpha, \beta $ represent $\pi N$ or $\gamma N$ channels.
All parameters $R^{(a)}_{\beta,\alpha}, B_{\beta,\alpha}^{(a)}$ and
$M_R^{(a)}$ are obtained numerically from the amplitude as described
in the previous section.
The above two amplitudes with $a=u,p$  can be combined by using the
following unified representation
\begin{eqnarray}
T_{\beta,\alpha}(p^{on}_\beta,p^{on}_\alpha, E \rightarrow M_R)
 & = & -\frac{R_{\beta,\alpha}+R^1_{\beta,\alpha}
p_{\pi\Delta} } {E - M_R - \gamma p_{\pi\Delta}}
 + B_{\beta,\alpha}
 + B^1_{\beta,\alpha} p_{\pi\Delta},  \label{eq:twochanex}
\end{eqnarray}
where $p_{\pi\Delta}$ is the $\pi\Delta$ on-shell momentum $p_x$
determined by Eq.(11).
We require $T_{\beta,\alpha}=T_{\beta,\alpha}^{(p/u)}$
at $p_{\pi\Delta}=p_{\pi\Delta}^{(p/u)}$.
This requirement for $\alpha=\pi N, \gamma N$ and $\beta=\pi N$ determines 
6 unknown  complex numbers $R,R^1,M_R,B,B^1$ and $\gamma$
from known parameters $R^{(a)}_{\beta,\alpha}, B_{\beta,\alpha}^{(a)}$
and $M_R^{(a)}$.
 Neglecting small contribution
of $R^1$ and $B^1$, we then obtain
\begin{eqnarray}
T_{\beta,\alpha}(p^{on}_\beta,p^{on}_\alpha, E \rightarrow M_R)
 & = & -\frac{R_{\beta,\alpha}}{E - M_R - \gamma
 p_{\pi\Delta}} + B_{\beta,\alpha} \label{eq:twochan}
\label{eq:res-2cbw}
\end{eqnarray}
where
\begin{eqnarray}
\gamma & = & \frac{M_{R}^{(p)} - M_{R}^{(u)}}
  {p_{\pi\Delta}^{(p)}- p_{\pi\Delta}^{(u)}}\label{eq:rr0}  \\
M_R & = & M_{R}^{(p)} - \gamma p_{\pi\Delta}^{(p)} \\
R^1_{\beta,\alpha} & = &
\frac{ R_{\beta,\alpha}^{(p)}(1 - \gamma dp_{\pi\Delta}^{(p)}/dE)
   -   R_{\beta,\alpha}^{(u)}(1 - \gamma dp_{\pi\Delta}^{(u)}/dE)}
  {p_{\pi\Delta}^{(p)}- p_{\pi\Delta}^{(u)}} \\
R_{\beta,\alpha} & = & R_{\beta,\alpha}^{(p)}(1 - \gamma
 dp_{\pi\Delta}^{(p)}/dE)   -  p_{\pi\Delta}^{(p)}R^1_{\beta,\alpha}.
\label{eq:rrr}
\end{eqnarray}
With $p^{(u)}_{\pi\Delta}=49 -68i$ MeV, $M^{(u)}_R=(1359 - 76i)$ MeV and
$p^{(p)}_{\pi\Delta}= -65+86i$ MeV, $M^{(p)}_R=(1357-76i)$ MeV, we have
$M_R = (1364 - 105i)$MeV, $\gamma= -0.146 + 0.062i$
and $R_{\pi N,\pi N}= (- 12 - 47i)$MeV. 
The quantities $R^{(u/p)}_{ \pi N, \gamma N}$ at $Q^2$ can be obtained from
$\bar{\Gamma}^R_{\pi N}\bar{\Gamma}^R_{\gamma N}$ of Eq.(\ref{eq:res-pole})
and hence $R_{ \pi N, \gamma N}$ can also be calculated from using
Eqs.(\ref{eq:rr0})-(\ref{eq:rrr}).
 By interpreting $R_{\pi N,\pi N}$ and $R_{\pi N,\gamma N}$
of Eq.(\ref{eq:res-2cbw})
as the residues of a pole 
and using the procedures described above,
 we can then extract the
electromagnetic helicity amplitudes $A_\lambda(Q^2)$ and $S_\lambda(Q^2)$.

\begin{figure}
\includegraphics[width=6cm]{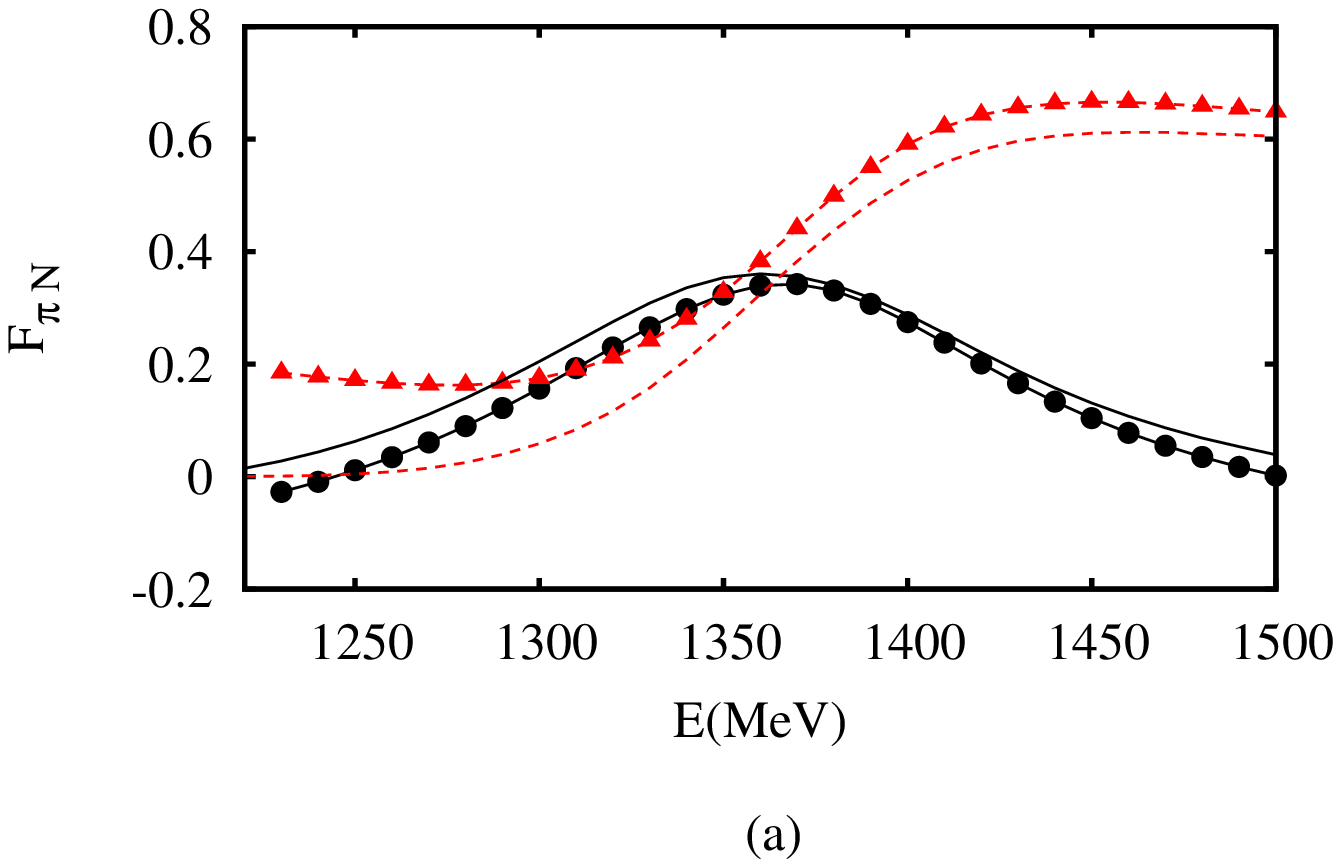} \hspace*{2cm}
\includegraphics[width=6cm]{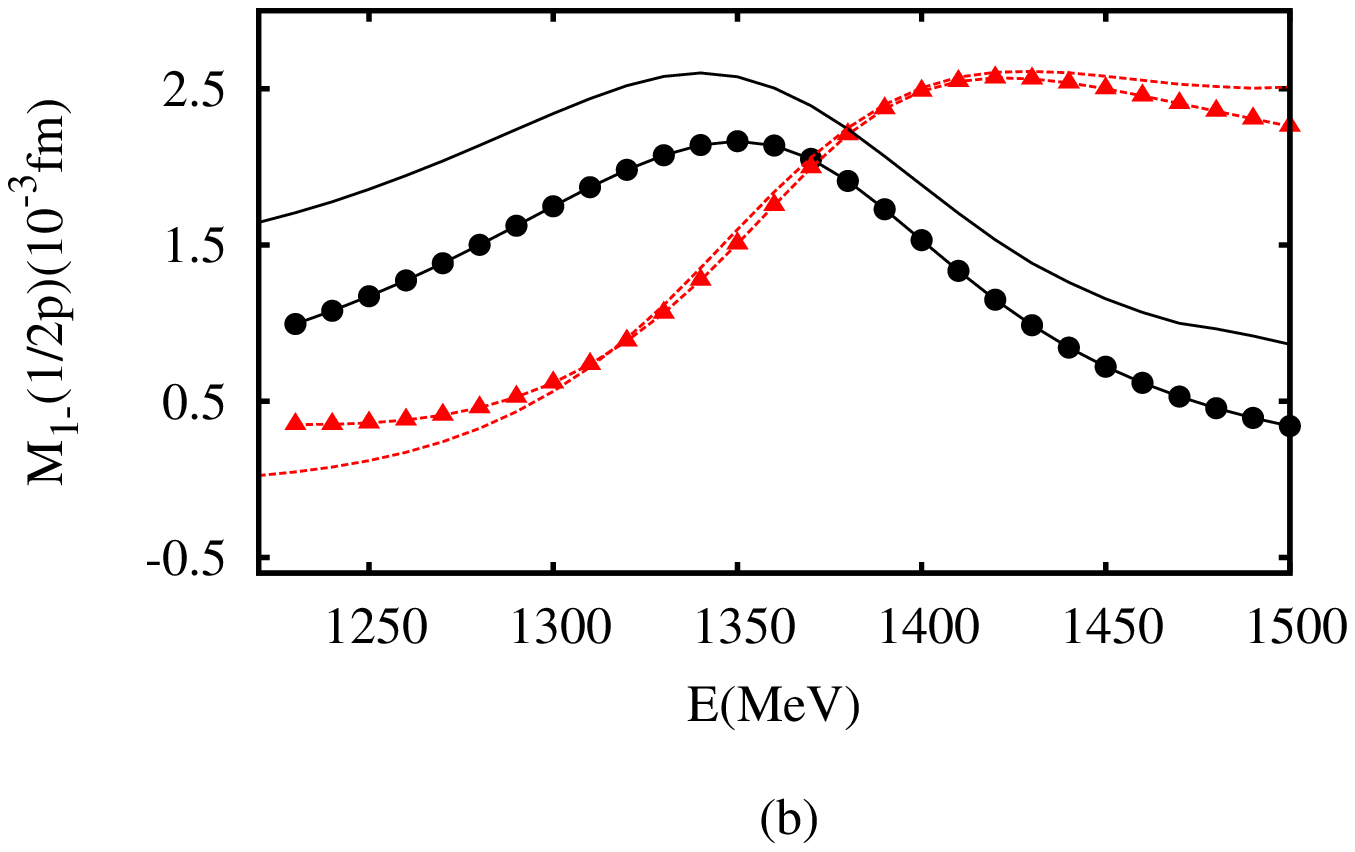}
\caption{Energy dependence of the $P_{11} \pi N$ scattering(a)
and the  $M_{1-}(1/2p)$ $\gamma\pi$ amplitude(b).
The solid circle(triangle) shows real(imaginary) part of the
amplitude calculated using Eq. (\ref{eq:twochan}).
The solid (dashed) curve shows real(imaginary) part of the amplitude
of EBC-DCC model.}
\label{fig:twochan-pin}
\end{figure}

We have found that the unified formula Eq. (\ref{eq:twochan}) 
 is a good approximation for both $\pi N$ and $\gamma\pi$ amplitudes
if  Eq.(44) is evaluated in the physical region where $E$ is near $W_R=Re(M_R)$.
This is shown in Fig. \ref{fig:twochan-pin} for
the considered $P_{11}$ partial wave.
 Although Eq.(\ref{eq:twochan}) is close to the commonly
used Breit-Wigner form of Eq.(\ref{eq:res-bw}), it is difficult to compare
the extracted $\gamma^* N \rightarrow N^*$ helicity amplitude
$A_\lambda(Q^2)$ with those from previous analysis using Breit-Wigner
parametrization, for the same reasons discussed above for the isolated
$P_{33}$ and $D_{13}$ resonances.

We now present  in Table \ref{tab:residues-gn} our results 
for the $\gamma^* N \rightarrow N^*$ for
the $P_{33}$, $D_{13}$ and $P_{11}$ resonances at 
$Q^2=0$ photon point. As comparisons, we also list
several previous results\cite{arndt04,ahrens04,dugger07,blanpied01}
which were extracted from using the
Briet-Wigner parametrization of resonant
amplitude.
 Our results are complex numbers, as expected 
from expression Eqs.(35)-(37). Here we mention that a recent 
 nucleon resonance
analysis\cite{bonn-catchina}  also yields complex
helicity amplitudes. 

We observe in Table \ref{tab:residues-gn} that the real parts
of our results for $P_{33}$ and $D_{13}$
 are in good agreement with
the listed previous results.
For the $P_{33}$ case, 
this good agreement is perhaps related to the fact 
that the imaginary parts of our
results for this pronounced resonance is much smaller than their real parts.
For $D_{13}$, a more detailed analysis is needed to
understand this comparison since $D_{13}$ involves large $\pi N$
inelasticity and our results have large imaginary parts.
For $P_{11}$ resonances, the real parts of
our results (2c-bw) calculated from using the unified
form Eq.(\ref{eq:res-2cbw}) do not agree with
the previous analysis using Breit-Wigner parametrization Eq.(\ref{eq:res-bw}).
This is perhaps also
related to the fact that our  results 
 for
each pole near $\pi\Delta$ threshold have large imaginary
parts, as also seen in Table \ref{tab:residues-gn}.

 \begin{table}[t]
\caption{The extracted $\gamma N \rightarrow N^*$ helicity amplitudes 
($A_{\lambda}$ in $10^{-3}GeV^{-1/2}$) are compared with previous results.}
 \begin{tabular}{|c|c|c|c|c|c|c|}
\hline
              &         & EBAC     & Arndt\cite{arndt04} & Ahrens\cite{ahrens04}  &  Dugger\cite{dugger07} & Blanpied\cite{blanpied01} \\
\hline
 $P_{33}(1210)$  & $A_{3/2}$ & -265+19i & $-243\pm 1$      &   $-256\pm
		  3$  &     $-258\pm 5$ & $-266.9\pm 1.6 \pm 7.8$ \\
                 & $A_{1/2}$ & -129+44i & $-129\pm 1$      &   $-137\pm
		  5$  & $-139\pm 4$  &  $-135.7 \pm 1.3 \pm 3.7$\\
\hline
 $D_{13}(1527)$  & $A_{3/2}$ & 171+91i & $167\pm 5$     &   $147\pm 10$ & 
$143\pm 2$ &\\
                 & $A_{1/2}$ & -31+29i  & $-20\pm 7$   &   $-38\pm 3 $ & 
$ -28\pm 2$ &\\
\hline
$P11(2cbw)$                   & $A_{1/2}$ & -28+20i & $-63\pm 5$        & &$-51\pm 2$ &\\
$\,\,\,\,\,\,\,\,(1356)$      & $A_{1/2}$ & -13+20i &     &    &  &\\
$\,\,\,\,\,\,\,\,(1364)$      & $A_{1/2}$ & -14+22i &     &    & & \\\hline
 \end{tabular}
\label{tab:residues-gn}
 \end{table}

\begin{figure}[b]
\centering
\includegraphics[width=8cm]{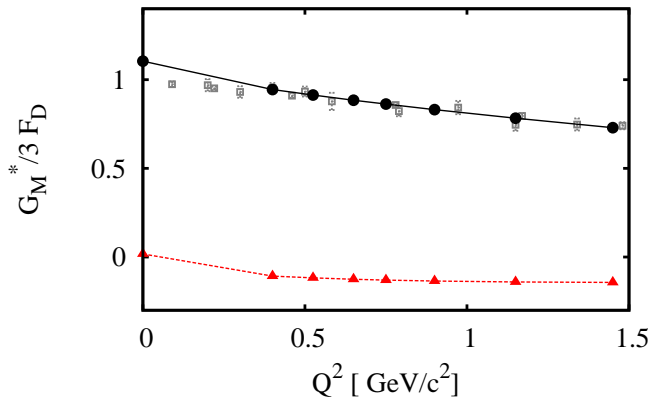}
\caption{The magnetic $N$-$\Delta$ (1232) transition form factor
$G^*_M(Q^2)$ defined in Ref.\cite{sl96}. $F_D=1/(1+Q^2/b^2)^2$ with
$b^2 = 0.71$ (GeV/c)$^2$. The solid circles (solid triangles) are 
the real (imaginary) parts of our results. The other data points are
 from previous analysis\cite{ndff-data}. }
\label{fig:p33-ff}
\end{figure}

For $P_{33}$ we can use the standard relation\cite{sl96} to
evaluate the $N$-$\Delta$ magnetic transition form factor $G_M^*$
in terms of helicity amplitudes. The real parts of
our results (solid circles connected by solid curve)
in Fig.\ref{fig:p33-ff} are in good agreement with the 
results (open circles with errors) from
the  previous analysis\cite{maid,inna}
using Breit-Wigner parametrization.  
In the same figure, we also show that the imaginary parts 
(triangles connected by dotted line)
of our results are much weaker. This observation further suggests
that our results could be close to
the results from analysis based on the
Breit-Wigner parametrization  only for
the cases that the imaginary parts of our results are small.

\begin{figure}[th]
\centering
\includegraphics[width=5cm]{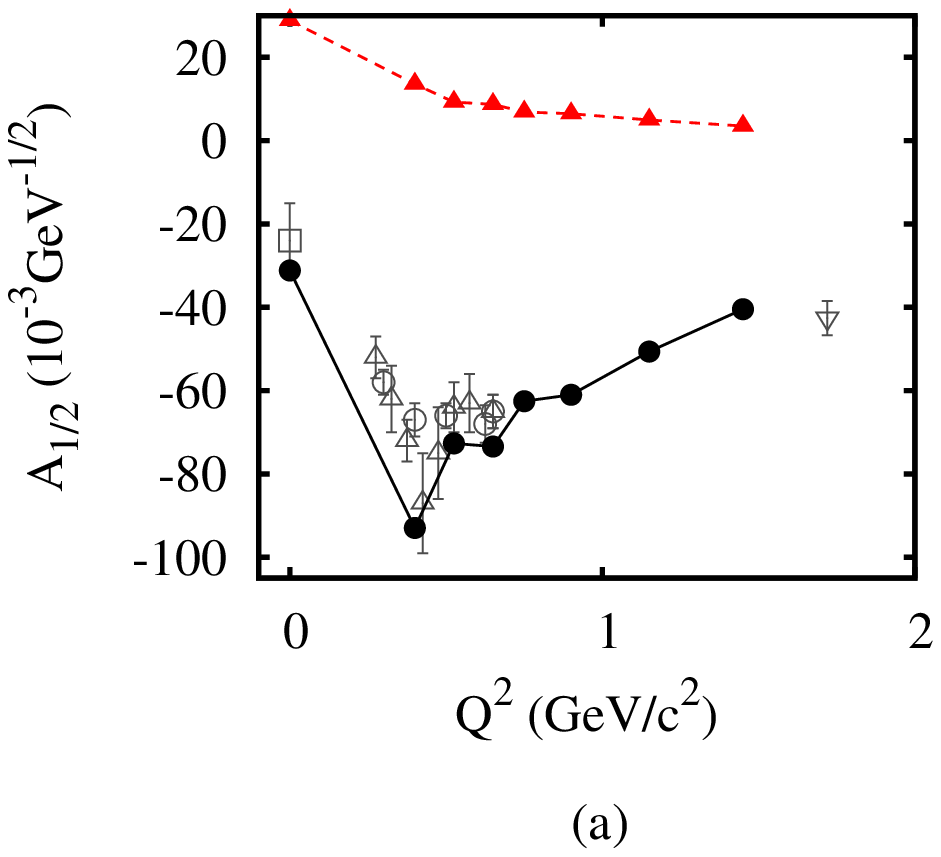}
\includegraphics[width=5cm]{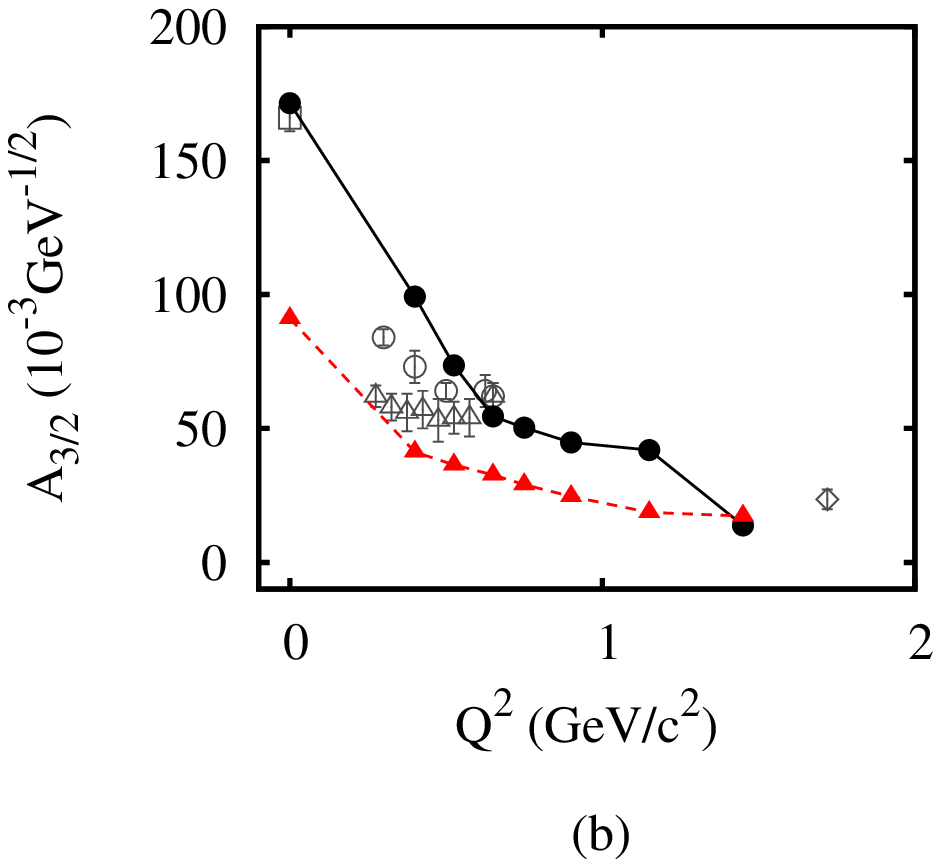}
\includegraphics[width=5cm]{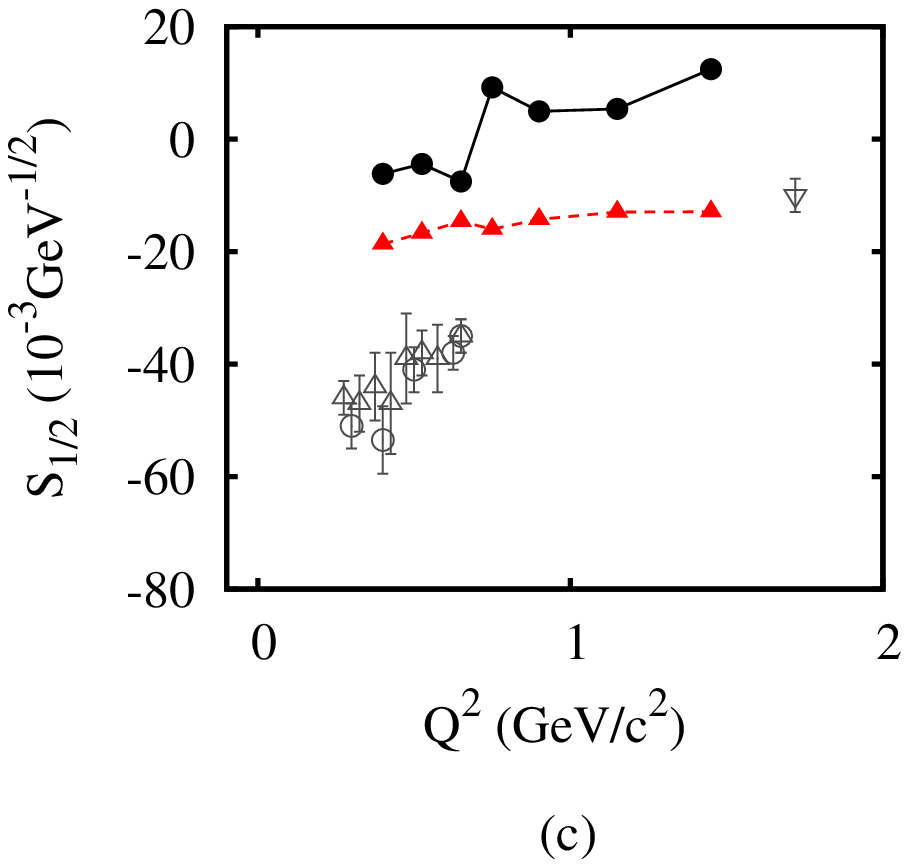}
\caption{
The extracted
 $\gamma N \rightarrow N^*(D_{13}(1527))$ form factors
($A_{1/2}$(a),$A_{3/2}$(b),$S_{1/2}$(c)).
The solid circles(solid triangles) are their the real(imaginary) parts.
The data are from CLAS collaboration\cite{inna}.}
\label{fig:d13-ff}
\end{figure}

For the  $D_{13}(1527)$ resonance, our results 
 are shown in Fig. \ref{fig:d13-ff}.
 As an example in seeing the difficulty in comparing our results
with those extracted from analysis using Breit-Wigner parametrization,
we also show the 
the results (open circles with errors) 
from CLAS collaboration\cite{inna}.
Qualitatively, CLAS analysis is based on the Eq.(\ref{eq:res-bw}) with
the choice of  $\phi^{BW}=0$. 
Thus their Breit-Wigner amplitude become
pure imaginary at $E = W_R$ with $W_R$ taken from PDG.
As discussed in the
beginning of this section, i.e. expression Eq.(17),  
the phase factor $e^{i\phi(E)}$ is
a necessary consequence of the presence of the non-resonance term
$B_{\pi N, \gamma^*N}(p,Q^2 ; E)$ under the unitarity condition.
This difference between the CLAS analysis and the previous
 analysis\cite{dalitz,taylor,mcvoy} 
 should be noted in interpreting their
extracted $\gamma^* N \rightarrow$ form factors. 

Despite the differences between two different analysis, 
we observe that
the real parts (solid circles connected by solid curve) of our 
 $A_{3/2}$ and $A_{1/2}$ shown in Fig. \ref{fig:d13-ff}
are qualitatively similar to the CLAS data. 
 The imaginary parts
(solid triangle connected by dashed curve) of our results,
which
are smaller than the real parts
but still appreciable, are also shown there.
Since the longitudinal parts of the amplitudes could not be well
determined with the available data, 
 the large
differences between our results and the CLAS data seen in
Fig. \ref{fig:d13-ff}
are not very surprising.

\begin{figure}[th]
\centering
\includegraphics[width=5cm]{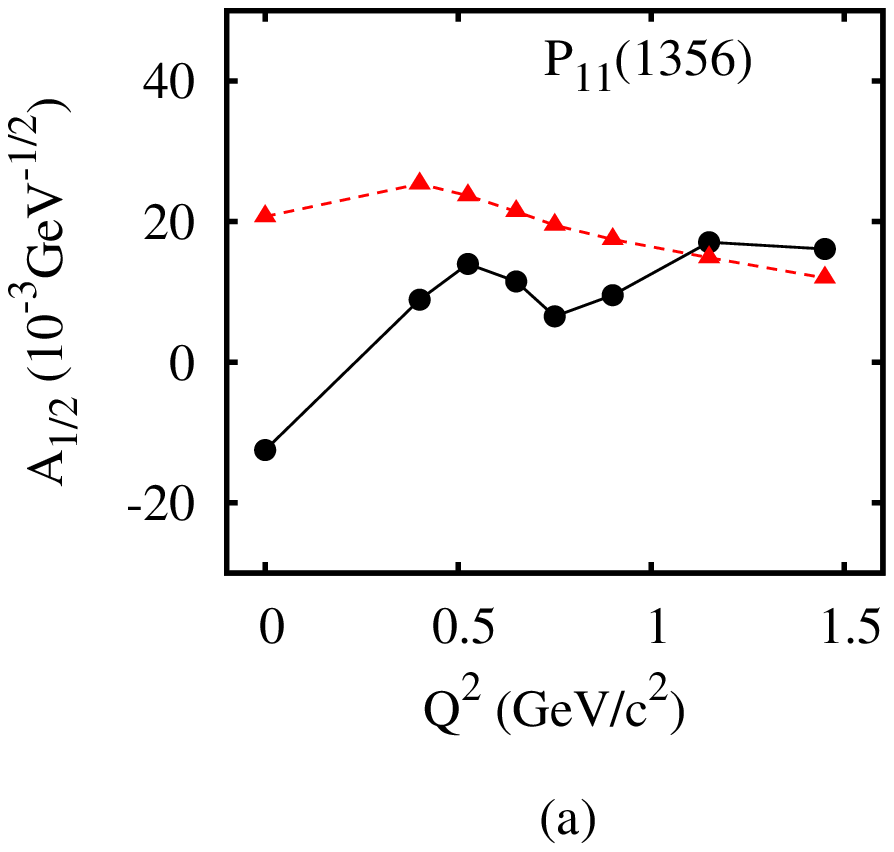}
\includegraphics[width=5cm]{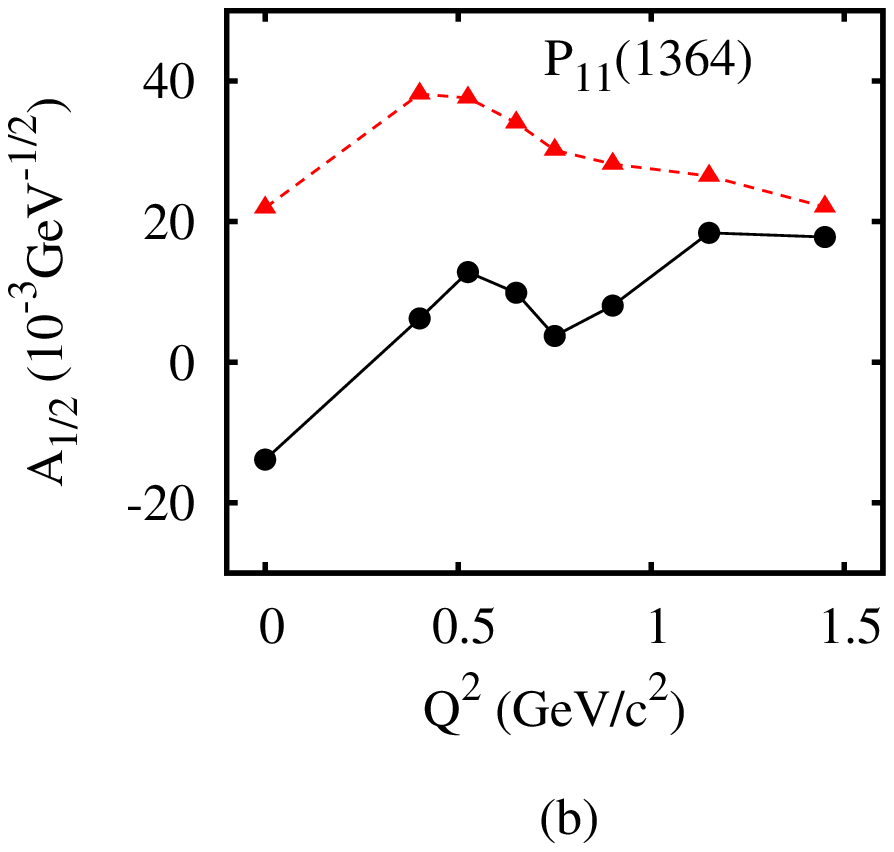}
\includegraphics[width=5cm]{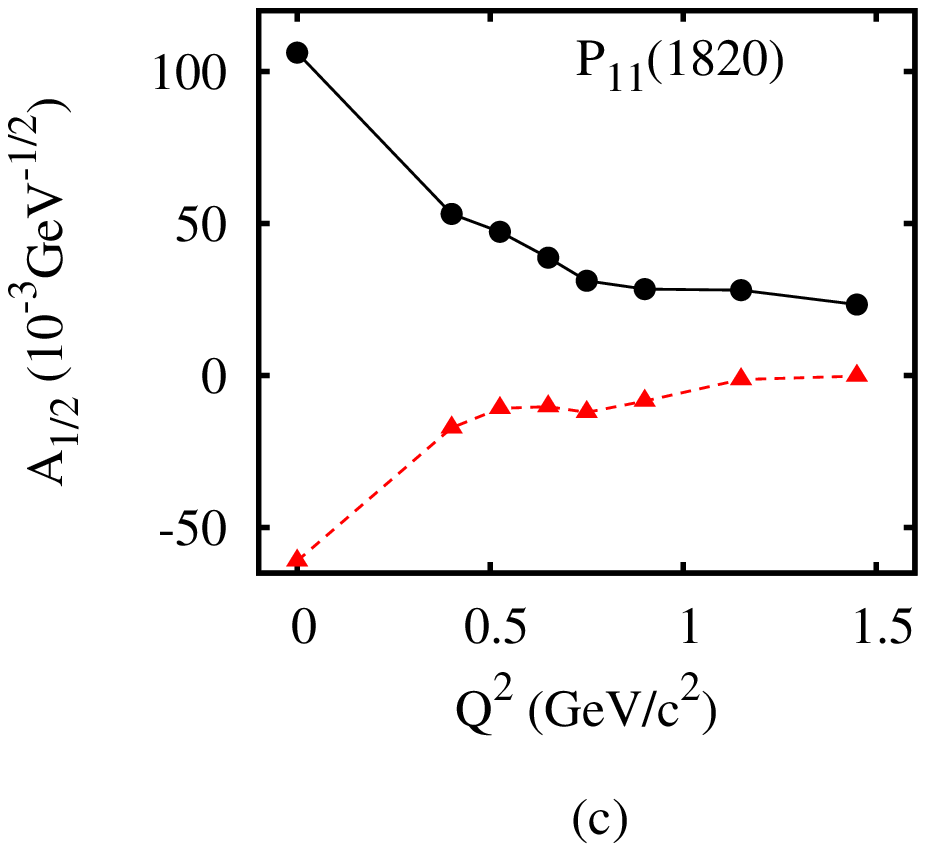}
\caption{
The extracted 
$\gamma N \rightarrow N^*(1356)$(a),$N^*(1364)$(b)
and $N^* (1820)$(c) transition form factors. 
The solid circles (solid triangles) are
their real (imaginary) parts.  }
\label{fig:p11-ff}
\end{figure}

Our results for the three poles
of $P_{11}$ listed in Table I are shown in Fig. \ref{fig:p11-ff}.
Similar to the results at photon point presented in Table III,
their imaginary parts (solid triangles) are  comparable
or larger than the real parts(solid circles) in magnitudes.
We note that the momentum dependence of the helicity amplitudes indicates
that the structure of $N^*(1356)$ and $N^*(1364)$ is
quite different from $N^*(1820)$.

\begin{figure}
\includegraphics[width=8cm]{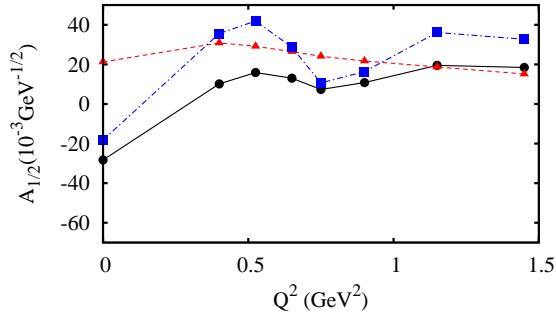}
\caption{  Helicity amplitude $A_{1/2p}$ of $P_{11}$ resonance
at $\pi\Delta$ threshold 
extracted from using the unified representation Eq.(46).
Solid (triangle) shows the real(imaginary) part of the
helicity amplitude. 
Solid square (connected by dot-dashed line) 
shows contribution of the bare form factor.}
\label{fig:twochan-em}
\end{figure}

 It is perhaps  more appropriate to interpret 
 our results calculated from using the unified
form Eq.(44) for the two poles
near the $\pi\Delta$ threshold as the values associated with the
Roper $N^*(1440)$ resonance listed by PDG.
This results are shown in Fig. \ref{fig:twochan-em}. We again see that its
imaginary parts (dotted line) are comparable or larger than
real parts (solid line) in most of the $Q^2$ region.
Here we also see that the
contribution ( dot-dashed lines)
from the determined bare $\gamma N \rightarrow N^*$ strengths play an
important role in changing the sign of the real part at $Q^2\sim $
0.4 (GeV/c)$^2$. This sign change of the bare 
$\gamma N \rightarrow N^*$ form factor is seen in some
relativistic constituent quark model 
calculations\cite{capstick,inna-1}.
This suggests that our bare parameters can perhaps be interpreted in terms of
hadron structure calculations excluding the  meson-baryon coupled-channel 
effects which is determined by unitarity condition.
Here we mention that our real parts are qualitatively similar to
the results from CLAS collaboration. But it is not clear how to
make connection between two results since we have very appreciable
imaginary parts.

\section{summary}
 
We have explained the application of a recently developed
 analytic continuation method to extract
the electromagnetic transition form factors for
the nucleon resonances ($N^*$)  within the EBAC-DCC model
of meson-baryon reactions. 
We discuss in detail how the contours for solving the
considered coupled-channels integral
equations are chosen to find resonance poles $M_R$  and their residues.
The formula for determining the $\gamma^* N \rightarrow N^*$ transition
form factors $A_\lambda(Q^2)$ and $S_\lambda(Q^2)$, defined
on the complex Riemann energy sheet, from the extracted
residues are presented.

We have found that the resulting Laurent expansions of the
$\pi N \rightarrow \pi N$ and $\gamma N \rightarrow\pi N$ amplitudes, evaluated
in the physical energy region, can reproduce to a very large extent
the exact solutions of EBAC-DCC model at energies near $E = Re(M_R)$.
A formula has been
developed to give an unified representation of the effects due to 
the first two $P_{11}$ resonances, which are near the $\pi\Delta$ threshold,
but are on different Riemann sheets.
Illustrative results 
for the well isolated  $P_{33}$ and $D_{13}$, and the complicated
$P_{11}$ resonances are presented.

We discuss  the differences between
our results and those extracted from the approaches using the Breit-Wigner
parametrization
of resonant amplitude to fit the data. 
We find that there is no simple
connection between   these two different
approaches,  despite that some of the real parts 
of our results and the results from Breit-Wigner analysis 
agree qualitatively when
the imaginary parts of our results are much smaller.

To conclude, we emphasize  that our form factors are defined
in a well-studied theoretical framework\cite{dalitz,taylor,mcvoy}
within which a resonance is an
"eigen state" of the Hamiltonian with
the outgoing boundary condition for the asymptotic wavefunction of its decay
channels. Thus the electromagnetic
transition form factors defined by $<\psi^R_{N^*}|J_{em}|N>$ , which can be
 extracted from the
residues of resonance poles, must be complex, since the
 resonant wavefunction  $\psi^R_{N^*}$ contains scattering continuum.
This must be accounted for in comparing our results with those
from using the Breit-Wigner form to fit the data and any hadron
structure calculations of $N$-$N^*$ transition form factors,
such as those from relativistic quark models\cite{capstick,inna-1},
Dyson-Schwinger models\cite{roberts}, and LQCD\cite{lqcd}.

\begin{acknowledgments}
This work is supported 
by the U.S. Department of Energy, Office of Nuclear Physics Division, under
Contract No. DE-AC02-06CH11357, and Contract No. DE-AC05-06OR23177
under which Jefferson Science Associates operates Jefferson Lab, and by
the Japan Society for the Promotion of Science,
Grant-in-Aid for Scientific Research(C) 20540270.
\end{acknowledgments}


\begin{thebibliography}{90}


\bibitem{ssl09}
N. Suzuki, T. Sato and T. -S. H, Lee,
Phys. Rev. C{\bf 79}, 025205 (2009).

\bibitem{sjklms10}
N. Suzuki, B. Julia-Diaz, H. Kamano, T.-S. H. Lee, A. Matsuyama, T. Sato,
Phys. Rev. Lett. {\bf 104}, 042302 (2010)

\bibitem{jlms07}
B. Julia-Diaz, T. -S. H. Lee, A. Matsuyama, and T. Sato,
Phys. Rev. C{\bf 76},  065201 (2007).

\bibitem{msl07}
A. Matsuyama, T. Sato, and T. -S. H. Lee,  Phys. Rept. {\bf 439}, 193 (2007).


\bibitem{afnan}
B.~C.~Pearce and I.~R.~Afnan,
Phys. Rev. C {\bf 34}, 991 (1986); {\bf 40}, 220 (1989).

\bibitem{gross}
F.~Gross and Y.~Surya,
Phys. Rev. C {\bf 47}, 703 (1993).

\bibitem{sl96}
T. Sato, T-.S. H. Lee, Phys. Rev. C{\bf 54}, 2660 (1996)

\bibitem{ntuanl}
C.~T.~Hung, S.~N.~Yang, and T.-S.~H.~Lee,
Phys. Rev. C {\bf 64}, 034309 (2001).

\bibitem{juelich-0}
A.~M.~Gasparyan, J.~Haidenbauer, C.~Hanhart, and J.~Speth,
Phys. Rev. C {\bf 68}, 045207 (2003);
M.~D\"oring, C.~Hanhart, F.~Huang, S.~Krewald, and U.-G.~Mei{\ss}ner,
Nucl. Phys. {\bf A829}, 170 (2009).


\bibitem{jlmss08}
B. Julia-Diaz, T. -S. H. Lee, A. Matsuyama, T. Sato and L. C. Smith,
Phys. Rev. C{\bf 77}, 045205 (2008).

\bibitem{jklmss09}
B. Julia-Diaz, H. Kamano, T. -S. H. Lee, A. Matsuyama, T. Sato
and N. Suzuki, Phys. Rev. C{\bf 80}, 025207 (2009).


\bibitem{bohm}
A. Bohm,
{\it Quantum mechanics: foundations and applications} 
(Springer-Verlag, New York, 1993).

\bibitem{dalitz} R. H. Dalitz and R. G. Moorhouse,
Proc. Roy. Soc. Lond. {\bf A318}, 279 (1970).



\bibitem{juelich}
M. D\"{o}ring, C. Hanhardt, F. Huang, S. Krewald and U.-G. Mei\ss ner,
Phys.Lett. {\bf B681}, 26 (2009)


\bibitem{gwu-vpi}
R. A. Arndt, W. J. Briscoe, I. I. Strakovsky, and R. L. Workman,
Phys. Rev C{\bf 74}, 45205 (2006).

\bibitem{maid}
D. Drechsel, S. S. Kamalov and L. Tiator,
Eur. Phys. J. A {\bf 34}, 69 (2007).

\bibitem{inna}
I.G. Aznauryan,V,D. Burkert, et al. (CLAS Collaboration),
Phys. Rev. C{\bf 80}, 055203 (2009);
V.I. Mokeev, V.D. Burkert, L. Elouadrhiri, G.V. Fedotov, E.N. Golovach,
and B.S. Ishkhanov, 
Chin. Phys. C{\bf 33}, 1210 (2009).

\bibitem{taylor}
R. J. Eden and J. R. Taylor, Phys. Rev. Lett. {\bf 11}, 516 (1963).

\bibitem{mp87}
D. Morgan and M.R. Pennington, Phys. Rev. Lett. {\bf 59}, 2818 (1987).


\bibitem{taylorbook}
J. Taylor, {\it Scattering Theory}
(Wiley, New York, 1972).

\bibitem{mcvoy}
K. W. McVoy, in 
{\it Fundamentals in Nuclear Theory}, edited by
A. De-Shalit and C. Villi(IAEA, Vienna, 1967), p475.


\bibitem{copley}
L. A. Copley, G. Karl, and E. Obryk
Nucl. Phys. {\bf B13},  303 (1969).


\bibitem{pdg}
C. Amsler et al., Phys. Lett. {\bf B667}, 1 (2008).

\bibitem{vpi84}
R.A. Arndt, J. M. Ford, L. D. Roper, Phys. Rev. D{\bf 32}, 1085 (1985).

\bibitem{cut}
R.E. Cutkosky and S. Wang, Phys. Rev. D. {\bf 42}, 235 (1990);
R. E. Cutkosky, C. P. Forsyth, R. E. Hendrick and R. L. Kelly,
Phys. Rev. D{\bf 20}, 2839 (1979).


\bibitem{kato65}
M. Kato, Ann. Phys. (N.Y.) {\bf 31}, 130 (1965).


\bibitem{fujii1}
Y. Fujii and M. Kato, Phys. Rev. {\bf 188}, 2319 (1969).

\bibitem{fujii2}
Y. Fujii and M. Fukugita, Nucl. Phys. {\bf B85}, 179 (1975).

\bibitem{bugg}
D. Bugg, J. Phys. G Nucl. Part. Phys. {\bf 37},  055002 (2010).

\bibitem{arndt04}
R. A. Arndt, W. J. Briscoe, I. I. Strakovsky, and
R. L. Workman, Phys. Rev. C{\bf 66}, 055213 (2002);
R. A. Arndt, I. I. Strakovsky, and R. L. Workman,
Phys. Rev. C{\bf 53}, 430 (1996).


\bibitem{ahrens04}
J. Ahrens et al., Eur. Phys. J. A{\bf 21}, 323 (2004);
J. Ahrens et al., Phys. Rev. Lett, {\bf 88}, 232002 (2002).

\bibitem{dugger07}
M. Dugger et al., Phys. Rev. C{\bf 76}, 025211 (2007).


\bibitem{blanpied01}
G. Blanpied et al., Phys. Rev. C{\bf 64}, 025203 (2001).

\bibitem{bonn-catchina}
A.V. Anisovich, E. Klempt,V.A. Nikonov, M.A. Matveev, A.V. Sarantsev,
and  U. Thoma,
Eur. Phys. J. A{\bf 44}, 203 (2010).

\bibitem{ndff-data}
W. Bartel et al.,Phys. Lett {\bf 28B}, 148 (1968);
K. B\"{a}tzner et al., Phys. Lett. {\bf 39B}, 575 (1972);
J. C. Alder et al., Nucl. Phys. {\bf B46}, 573  (1972);
S. Sterin et al., Phys. Rev. D{\bf 12}, 1884 (1975).

\bibitem{capstick}
S. Capstick and B.D. Keister, Phys. Rev. {\bf D51}, 3598 (1995)  
 
\bibitem{inna-1}
I.G. Aznauryan, Phys. Rev. {\bf C76}, 025212 (2007)

\bibitem{roberts}
See the review by P. Maris and C.D. Roberts,
Int.J.Mod.Phys. {\bf E12} 297(2003).

\bibitem{lqcd}
H-W. Lin et al., Phys. Rev. {\bf D79}, 034502 (2009).

\end{thebibliography}
\end{document}